\documentclass[aps,prd,longbibliography,groupedaddress,notitlepage,nofootinbib,11pt]{revtex4-2}
\usepackage{amsmath}
\usepackage{amssymb}
\usepackage{natbib}
\usepackage{url}
\usepackage{float}
\usepackage{graphicx,subfigure}
\usepackage{hyperref}
\usepackage{braket}
\usepackage{comment}
\usepackage[normalem]{ulem}

\newcommand{\be}{\begin{equation}}
 \newcommand{\ee}{\end{equation}}
 \newcommand{\bse}{\begin{subequations}}
 \newcommand{\ese}{\end{subequations}}
\newcommand{\ba}{\begin{eqnarray}}
\newcommand{\ea}{\end{eqnarray}}

\usepackage{mathtools}
\usepackage[framemethod=tikz]{mdframed}
\usepackage{tikz}
\usetikzlibrary{decorations.pathreplacing,decorations.markings}
\usepackage{subfigure}
\usepackage{comment}
\usetikzlibrary{shapes.geometric, arrows}
\tikzstyle{startstop} = [rectangle, rounded corners, minimum width=2cm, minimum height=1cm,text centered, text width=4.4cm, draw=black, fill=white!30]
\tikzstyle{io} = [Trapezium, Trapezium left angle=70, Trapezium right angle=110, minimum width=2cm, minimum height=1cm, text centered, draw=black]
\tikzstyle{process} = [rectangle, minimum width=1.5cm, minimum height=1cm, text centered,, text width=4cm, draw=black, fill=white!30]
\tikzstyle{decision} = [diamond, minimum width=3cm, minimum height=1cm, text centered, text width=4.2cm, draw=black, fill=green!30]
\tikzstyle{arrow} = [thick,->,>=stealth]
\definecolor{mycolor}{rgb}{0.122, 0.435, 0.698}
\newmdenv[innerlinewidth=0.5pt, roundcorner=4pt,linecolor=mycolor,innerleftmargin=6pt,
innerrightmargin=6pt,innertopmargin=6pt,innerbottommargin=6pt]{mybox}

\begin{document}


\title{A stabilizer code model with non-invertible symmetries: Strange fractons, confinement, and non-commutative and non-Abelian fusion rules}

\author{Tanay Kibe,}
\affiliation{Center for Operator Algebras, Geometry, Matter and Spacetime, Indian Institute of Technology Madras, Chennai $600036$, India}
\email{tanaykibe2.71@gmail.com}
\author{Ayan Mukhopadhyay,}
\affiliation{Instituto de F\'{\i}sica,
Pontificia Universidad Cat\'{o}lica de Valpara\'{\i}so,
Avenida Universidad 330, Valpara\'{\i}so, Chile.}
\altaffiliation[Also at ]{Center for Operator Algebras, Geometry, Matter and Spacetime, Indian Institute of Technology Madras, Chennai $600036$, India}
\email{ayan.mukhopadhyay@pucv.cl}

\author{Pramod Padmanabhan}
\affiliation{School of Basic Sciences, Indian Institute of Technology, Bhubaneswar, $752050$, India}
\email{pramod23phys@gmail.com}


\begin{abstract}

We introduce a stabilizer code model with a qutrit at every edge on a square lattice and with non-invertible plaquette operators. The degeneracy of the ground state is topological as in the toric code, and it also has the usual deconfined excitations consisting of pairs of electric and magnetic charges. However, there are novel types of confined fractonic excitations composed of a cluster of adjacent faces with vanishing flux. They manifest confinement, and even larger configurations of these fractons are fully immobile although they acquire emergent internal degrees of freedom. Deconfined excitations change their nature in presence of these fractonic defects. As for instance, fractonic defects can absorb magnetic charges making magnetic monopoles exist while electric charges acquire restricted mobility. Furthermore, some generalized symmetries can annihilate any ground state and also the full sector of fully mobile excitations. All these properties can be captured via a novel type of \textit{non-commutative} and \textit{non-Abelian} fusion category in which the product is associative but does not commute, and can be expressed as a sum of (operator) equivalence classes. Generalized non-invertible symmetries give rise to the feature that the fusion products form a non-unital category without a proper identity. We show that a variant of this model features a deconfined fracton liquid phase and a phase where the dual (magnetic) strings have condensed.
\end{abstract}

\maketitle
\section{Introduction and Summary of Results}
Our understanding of symmetries of physical systems has recently undergone a profound transformation \cite{Gaiotto:2014kfa,McGreevy:2022oyu,Schafer-Nameki:2023jdn,Shao:2023gho}. Firstly, the notion of symmetries has been extended by studying their action not only on local operators but also on non-local operators and defects. Secondly, it has been understood that such generalized symmetry operations are not necessarily invertible. Non-invertible generalized symmetries have now been found ubiquitously, as for instance, even in the standard model of particle physics after the inclusion of topological defects \cite{Choi:2022jqy,Putrov:2023jqi,Cordova:2022qtz}, and also in many condensed matter systems such as the Ising model \cite{McGreevy:2022oyu,Schafer-Nameki:2023jdn,Seiberg:2023cdc}.  

In a parallel development, a new class of quasiparticles has been discovered in a variety of systems which are fully immobile or quasi-immobile (implying that they have mobility only along sub-manifolds) \cite{Nandkishore:2018sel,Pretko:2020cko}. A typical feature of such systems featuring fractonic excitations is that the degeneracy of ground states scale with the size of the system and can also depend on the details of the lattice. These exotic fractonic quasiparticles have been partially integrated into the quantum field theory framework via symmetric tensor gauge theories \cite{xu2006novel,Xu2} and its variants which explains the (complete or partial) immobility of the fractons in terms of higher moment symmetries \cite{PretkoFT1,PretkoFT2,Seiberg:2019vrp,Bulmash:2018knk,HanHermeleXie,Gromov,Seiberg:20201,Seiberg:20202,Seiberg:20203} (see also \cite{FractonElasticity1,FractonElasticity2}). {Another approach for understanding systems with fractonic excitations involves gauging subsystem symmetries \cite{Xcube,Kim2,FractonSubsystem1,FractonSubsystem2,FractonSubsystem3,FractonSubsystem4} (more on this in Sec. \ref{Sec:Model}). Whether phases with fractonic excitations can be defined via generalized topological order has also been investigated \cite{Haahcode,Kim1,Aasen:2020zru}\footnote{See \cite{Kim1} particularly for the study of the X-cube model \cite{Xcube} where the notion of fractonic topological order is defined via the topology of singular compact total foliation of the underlying three manifold. See also \cite{Aasen:2020zru} for an understanding of fractons in terms of topological defect networks.}.}

A natural question, therefore is, whether one can find models with non-invertible (generalized) symmetries in which fractonic quasiparticles exist, and whether such models where non-invertible symmetries have a natural action on fractonic quasiparticles can be incorporated into the quantum field theory framework.
A natural place to look for such solvable models are stabilizer codes in which the Hamiltonian can be written in terms of elementary plaquette and dual plaquette operators (equivalently vertex operators), called elementary stabilizers, all of which commute with each other \cite{Kitaev:1997wr,Gottesman:1997zz}. In fact, stabilizer codes can naturally feature fractons \cite{Haahcode,Xcube}. Furthermore, the phases of stabilizer codes along with the quasiparticles and their fusion rules can be often incorporated into the gauge theory framework \cite{FradkinShenker,baez2005higher, ibieta2020topological, de2017topological, bullivant2017topological, bullivant2020higher,Barkeshli:2022edm,Pace:2022cnh}. One can hope that stabilizer code models featuring fractons and non-invertible symmetries, can be incorporated into the quantum field theory framework  leading to novel quantum field theories with wide ranging applications. 

{In this work, we construct and study a stabilizer code model with inbuilt non-invertible symmetries exhibiting novel type of fully immobile fractonic excitations. Particularly, we show that such a model with non-invertible symmetries and fractonic excitations necessitates the construction of novel non-commutative and non-Abelian fusion rules to characterize the excitations.\footnote{In this work, we will be interested not in the fusion category of only the symmetries but rather that of the superselection sectors of excitations of the theory, e.g. anyon fusion rules \cite{Kitaev:1997wr,nayak2008non}. The latter is similar to the fusion rules in conformal field theories \cite{Fuchs:1993et} which arise as a consequence of operator product expansion of primary operators. In our construction, the symmetries which are \textit{logical operators} close non-trivially under fusion rules, and some of these logical operators are non-invertible. However, we will avoid these logical operators for simplicity by localizing excitations on a finite sub-region of the lattice.} In order to construct fusion rules, we introduce the notion of spectral monoid, which is the set of operators which are closed under multiplication and with minimal number of generators such that they can generate \textit{all} local excitations by acting on any ground state (see Sec. \ref{Sec:Fusion} for a precise definition). We also define equivalence classes of operators in the spectral monoid which accommodates non-invertible operators appropriately, and utilize this to obtain the fusion rules which are associative but non-commutative and non-Abelian.  In fact, the construction of the spectral monoid and the fusion rules characterizes which equivalence classes are related to immobile fractonic excitations. It will be interesting to see how such a fusion category can be incorporated into a quantum field theory.}

Particularly, we construct a simple and exactly solvable stabilizer code model with a qutrit ($3$ level system) on each edge of a  two-dimensional lattice and which naturally generalizes the toric code (with a qubit on each edge) such that some of the elementary stabilizers are non-invertible. Thus the model has a strikingly intrinsic non-invertible symmetry, in the sense that it is built into the Hamiltonian of the theory itself. {In this model, we will show that the generalized symmetries are of two types: (i) those which keep any ground state invariant forming the \textit{stabilizer monoid}, and (ii) logical operators which have non-trivial actions on the ground states. Additionally there are Hermitian operators which commute with the  Hamiltonian and annihilate the ground states (and also the sector with only mobile excitations) forming the \textit{annihilator monoid}}. In our model, the stabilizer monoid is not generated simply by the elementary stabilizers, but also by non-invertible operators supported on contractible loops of arbitrary sizes (and therefore similar to generalized symmetries). Such symmetries have a natural action on the fractonic excitations, and in fact {the stabilizer monoid can completely distinguish all local excitations, including fractonic excitations and their internal states.}

When restricted to the full sector of states that can decay to a ground state in the presence of local perturbations, the symmetries, which form the stabilizer monoid and the set of logical operators, act as unitary transformations (representing the action of $\mathbb{Z}_2$), while the annihilator monoid annihilates this entire sector of states. However, outside of this sector, even the stabilizer monoid and the logical operators are not necessarily invertible.

Interestingly, although our model has fractonic excitations, the degeneracy of the ground states is exactly like the toric code, and is therefore determined only by the genus of the lattice and not its size. In the concluding section, we will introduce a general family of such models with non-invertible symmetries in which this need not be the case (the ground state degeneracy can scale with the system size and there exists fractonic excitations).  Nevertheless, it is worth mentioning that our model also has whole superselection sectors with deconfined mobile excitations that can decay to each ground state in presence of local perturbations. The fractonic excitations form new superselection sectors with finite energy barrier and therefore they are reminiscent of dark matter if the mobile excitations which can decay to the ground state are analogous to visible matter in the Universe.\footnote{This is only a provocative analogy. Here we are not claiming that we have a new theory for dark matter.} These mobile excitations are exactly like those in the toric code, consisting of pairs of deconfined electric or magnetic charges (these are $\mathbb{Z}_2$ gauge theory charges).

The primary fractonic excitations of our model are essentially connected clusters of face excitations with vanishing fluxes and in which each face shares at least one edge with another such face. Their immobility is rather extreme in the sense that they cannot be translated by any local operation, even accounting for an arbitrary energy cost.\footnote{In this aspect, the fractons in our model are partially similar to the elementary fractons in \cite{Yan:2018nco}. However, in the latter model, fracton bound states are mobile.} Furthermore, the zero flux face can be separated from the cluster only at the cost of creating other such face excitations, implying that the energy of the cluster grows linearly with the size. The latter is the hallmark of confinement. Furthermore, larger fractonic clusters are also completely immobile, but interestingly they acquire emergent internal degrees of freedom. Our fractons are not of type I \cite{Channon,bravyi2011topological,HaahFuVijay,Xcube}, as larger configurations of fractons do not gain even partial mobility although they gain novel internal degrees of freedom.

Remarkably, the deconfined electric and magnetic charges change their nature in presence of fractonic configurations. Particularly, a magnetic monopole can exist outside a fractonic cluster although it cannot exist otherwise. This magnetic monopole is also locally mobile. However, it cannot decay to a ground state in presence of local perturbations. Since some excitations which are not localized cannot also decay to the ground state, our fracton model is not of type II category \cite{Haahcode,Yoshida} as well.\footnote{A stricter notion of type II class is that all excitations of the system are completely immobile. Here we are employing the notion of type II class discussed in \cite{Pretko:2020cko}.} The deconfined electric charges acquire restricted mobility in the presence of fractonic defects as they cannot penetrate through them without annihilating the state (similar to the case of electrons obeying the Pauli exclusion principle).

As indicated above, we show that all the physical properties of the model can be characterized by novel fusion rules, which require the zero operator for closure, and which are both non-commutative and non-Abelian. The construction of the fusion rules will require us to define operator equivalence classes in a novel way. {The most striking aspect of the fusion rules is that fusion products of fractonic defect superselection sectors with the identity sector can be both non-commutative and non-Abelian reflecting the fact that non-invertible symmetries generally leads to non-trivial fusion products. Furthermore, such non-trivial products are necessary for the fusion rules to be associative.} 

In Sec. \ref{Sec:Decon}, we discuss a variant of our model in which there exsits two novel phases. One in which the fractonic excitations have deconfined to create a fracton liquid phase (analogous to a spin liquid), and another in which the dual (magnetic) strings have condensed.

The paper is organized as follows. In Section \ref{Sec:Model}, we introduce the model, prove that its ground state degeneracy is determined by the genus of the lattice, and discuss the generalized symmetries. In Section \ref{Sec:Excitations}, we discuss the deconfined excited states, the confined fractonic defects and how the deconfined fractonic excited states change their nature in the presence of fractonic defect configurations. Furthermore. we show that the commutative subset of the generalized symmetries of our model can distinguish (detect) all the local excitations of our model. We also discuss a variant of our model with a quantum phase in which the fractons can deconfine. In Section \ref{Sec:Fusion}, we construct the fusion rules of the model, which involve the zero operator for closure, and which are both non-commutative and non-Abelian. We also discuss how the fusion rules capture the physical nature of the excitations. In Section \ref{Sec:Outro}, we conclude by introducing more variants of our model and discussing the relevance of our work in extending the quantum field theory framework.

\section{The model}\label{Sec:Model}

A stabilizer code model is a quantum system defined on a lattice with a Hamiltonian that is a sum of local commuting operators, a.k.a. elementary stabilizers. Here we introduce such a model where each edge on a square lattice\footnote{The model is well defined on an arbitrary tiling of a Riemann surface with or without boundaries. The square lattice is used for simplicity and it sufficiently describes the system's features.} hosts a qutrit ($\mathbb{C}^{3}$) generalizing the $\mathbb{Z}_2$-toric code introduced in \cite{Kitaev:1997wr} where each edge hosts a qubit ($\mathbb{C}^{2}$). 

Let us define, in the Pauli $Z$ basis, the following Hermitian operators $M_{1,2}$ that act on this qutrit Hilbert space:
\begin{equation}
    M_1 \equiv 
    \begin{pmatrix}
        -1&0&0\\
        0&0&1\\
        0&1&0
    \end{pmatrix} = (-1) \bigoplus X, \quad
    M_2 \equiv 
    \begin{pmatrix}
        0&0&0\\
        0&1&0\\
        0&0&-1
    \end{pmatrix} = 0 \bigoplus Z,
\end{equation}
where $Z$ and $X$ are Pauli matrices. $M_2$ has a zero eigenvalue and is therefore a non-invertible  operator whereas $M_1$ is both Hermitian and unitary with eigenvalues $\pm 1$ (the $-1$ eigenvalue is doubly degenerate). It is easy to see that $M_1$ and $M_2$ anti-commute. Note also that $M_1^2 =1$ and $M_2^3 =M_2$.

It is important that $M_2$ is non-invertible and has a zero eigenvalue. Otherwise, $M_2$ would not anti-commute with $M_1$ (if $M_2$ is chosen to be $1 \bigoplus Z$, then it does not anti-commute with $M_1$.) 

The stabilizers for this code are the vertex operators $A_v$ attached to each vertex $v$, and plaquette operators $B_f$ attached to each face $f$, which are defined as follows:

\begin{equation}
A_{v} = \vcenter{\hbox{
\begin{tikzpicture}[scale=1.2]
\draw[black, thick] (0,1) -- (0,3);
\draw[black, thick] (-1,2) -- (1,2);
\filldraw[red] (0,1.5) circle (2pt) node[anchor=west]{\textcolor{black}{$M_1$}};
\filldraw[red] (0,2.5) circle (2pt) node[anchor=east]{\textcolor{black}  {$M_1$}};
\filldraw[red] (-0.5,2) circle (2pt) node[anchor=north]{\textcolor{black}{$M_1$}};
\filldraw[red] (0.5,2) circle (2pt) node[anchor=south]{\textcolor{black}{$M_1$}};
\filldraw[white] (0.1,2.1) circle (2pt) node[anchor=center]{\textcolor{black}{v}};
\end{tikzpicture}}}, \quad
    B_{f}= \vcenter{\hbox{
\begin{tikzpicture}[scale=0.55]
\draw[black, thick] (0,0) -- (0,4);
\draw[black, thick] (0,0) -- (4,0);
\draw[black, thick] (4,0) -- (4,4);
\draw[black, thick] (0,4) -- (4,4);
\filldraw[red] (0,2) circle (4pt) node[anchor=west]{\textcolor{black}{$M_2$}};
\filldraw[red] (4,2) circle (4pt) node[anchor=east]{\textcolor{black}{$M_2$}};
\filldraw[red] (2,0) circle (4pt) node[anchor=north]{\textcolor{black}{$M_2$}};
\filldraw[red] (2,4) circle (4pt) node[anchor=south]{\textcolor{black}{$M_2$}};
\filldraw[white] (2,2) circle (4pt) node[anchor=center]{\textcolor{black}{f}};
\end{tikzpicture}}}.
\end{equation}
Since $A_v$ and $B_f$ either do not overlap or overlap on two edges, they commute as $M_1$ and $M_2$ anti-commute. Thus 
\begin{equation}
    [A_{v}, A_{v'}]=  [B_{f}, B_{f'}]=  [A_{v}, B_{f}]=0.
\end{equation}
The Hamiltonian is the sum of these mutually commuting stabilizers:
\begin{equation}
    H = - \sum_v A_v -\sum_f B_f.
    \label{Eq:ham}
\end{equation}
The set of operators generated by products of $\{A_v\}$ and $\{B_f\}$ thus form a commutative monoid which we will denote as $S$. However, this is not the full stabilizer commutative monoid consisting of symmetry operations under which any ground state is left invariant, as will be discussed soon. Restriction of our model to the lower two-dimensional subspace on each edge (on which $M_1$ acts as $X$ and $M_2$ acts as $Z$) gives the $\mathbb{Z}_2$-toric code \cite{Kitaev:1997wr}.

We can view $B_f$ as the plaquette operator of $\mathbb{Z}_2$ gauge theory where the gauge transformation acts only on a subspace of the Hilbert space at each edge corresponding to the lower $2\times 2$ block where $M_2$ is non-vanishing. However, this feature is enforced by the requirement that $M_2$ anti-commutes with $M_1$. Our model has similarities to many models in which gauging subsystem symmetries leads to fractons \cite{Nandkishore:2018sel,Pretko:2020cko} and/or non-invertible defects \cite{Shao:2023gho}.

\subsection{Ground states have topological degeneracy}  A ground state of this system will be the simultaneous eigenstate of all $A_v$ and $B_f$ with the highest possible eigenvalues, namely $+1$, so that the energy is minimized. Thus any ground state can be written in the form:
\be
\ket{G} = \mathcal{P}_0 \ket{s}
\ee
where $\ket{s}$ is a suitable seed state and $\mathcal{P}_0$ is the product of projectors $P_v$ and $P_f$, which projects to the $+1$ eigenstate sector of $A_v$ and $B_f$, respectively, i.e.
\be\label{Eq:P0gs}
P_v = \frac{1}{2}(1 + A_v), \,\, P_f = \frac{1}{2}(B_f + B_f^2),\,\, \mathcal{P}_0 = \prod_v P_v\prod_f P_f.
\ee
By construction, $\mathcal{P}_0$ is the projector to the subspace spanned by \textit{all} groundstates. The degeneracy of the ground states (GSD) can be obtained simply by computing ${\rm Tr} (\mathcal{P}_0)$, i.e. the rank of $\mathcal{P}_0$. This can be readily done as ${\rm Tr}(M_2) = {\rm Tr}(M_1 M_2) = {\rm Tr}(M_1 M_2^2) =0$ implies that
\begin{multline}
   \mathcal{P}_0 = \frac{1}{2^{|v|} 2^{|f|}} \prod_f B_f +\frac{1}{2^{|v|} 2^{|f|}} \prod_f B_f^2 + \frac{1}{2^{|v|} 2^{|f|}}\prod_v A_v \prod_f B_f +\frac{1}{2^{|v|} 2^{|f|}} \prod_v A_v \prod_f B_f^2 \\+ \text{traceless terms} ,
\end{multline}
where $|v|$ and $|f|$ denote the number of vertices and faces on the lattice. Assuming that the lattice is a tessellation of a compact surface, we readily obtain $\prod_v A_v = I$ which further implies that
\begin{align}
    \mathcal{P}_0 = \frac{2}{2^{|v|} 2^{|f|}} \prod_f B_f +\frac{2}{2^{|v|} 2^{|f|}} \prod_f B_f^2 + \text{traceless terms} .
\end{align}
Both the terms $\prod_f B_f$ and $\prod_f B_f^2$ correspond to $M_2^2$ acting on each edge of the lattice (since $M_2^4 = M_2^2$). Therefore,
\begin{equation}
    \mathcal{P}_0 = \frac{4}{2^{|v|} 2^{|f|}} \prod_e M_2^2 + \text{traceless terms}.
\end{equation}
Since ${\rm Tr}M_2^2 = 2$, we finally obtain (with $|e|$ denoting the number of edges of the lattice):
\begin{equation}
    {\rm GSD}={\rm Tr}(\mathcal{P}_0) = {\rm Tr}\left(4\frac{1}{2^{|v|} 2^{|f|}} \prod_e M_2^2 \right) = 4\frac{2^{|e|}}{2^{|v|} 2^{|f|}} = 2^{2 +|e|- |v|-|f|} = 2^{2-\chi} = 2^{2g}.
\end{equation}
Above we have used the standard expression for the Euler characteristic of the closed surface $\chi = |v|+|f|-|e|$ in terms of its genus $g$, namely $\chi = 2 - 2g$. The model has therefore the same GSD as that of the toric code (where a similar but much shorter trace calculation yields the same answer elegantly \cite{ferreira20142d}).

\subsection{Symmetries: stabilizer monoid, annihilator monoid and logical operators}\label{Sec:GenSym}

In stabilizer codes, one usually identifies the stabilizer group, which is an Abelian group of Hermitian operators which commute with the Hamiltonian (and mutually commute with each other), and whose eigenvalues can be used to distinguish all locally excited states, i.e. states which are indistinguishable from a ground state outside of a finite region. Furthermore, the action of the stabilizer group leaves any ground state invariant \cite{Gottesman:1997zz}. In the context of the toric code, each element of the stabilizer group is not only Hermitian but also an unitary operator that commutes with the Hamiltonian. Therefore, each element of the stabilizer group can be viewed as a \textit{symmetry transformation} which leaves any ground state invariant.  The stabilizer group of the toric code is generated by the elementary stabilizers, i.e. the elementary vertex and plaquette operators which appear in the Hamiltonian.

Additionally, there are \textit{logical operators} in stabilizer codes which commute with the Hamiltonian but which can generate other ground states when they act on a specific ground state. These \textit{logical operators} do not necessarily commute with each other. In the toric code, such operators are also both unitary and Hermitian, and can be viewed as symmetry transformations which have non-trivial action on the manifold of ground states. 

A more convenient point of view for the toric code is to consider the set of all unitary transformations which leave the Hamiltonian invariant, or equivalently all unitary operators which commute with the Hamiltonian. The set of such operators form the group of symmetry transformations. The maximal Abelian subgroup which leaves the ground state invariant is the stabilizer group. The latter is generated by the elementary vertex and plaquette operators (elementary stabilizers). Taking the quotient of the group of symmetry transformations with the stabilizer group yields the logical operators which are commensurate with the number of ground states. 

\paragraph{\textbf{Generalized symmetries of our model}:} The notion of symmetries should be generalized in our model as some of the elementary stabilizers, especially the elementary plaquette operators $B_f$, are not invertible. In this context, it is useful to recall the case of the transverse field Ising model \cite{Aasen:2016dop,PhysRevB.91.195143,Shao:2023gho} where there exists an operator $\mathcal{D}$ which can be written as a product of an unitary operator $U$ and a projector $P_+$ (to the $\mathbb{Z}_2$ invariant subspace) so that $\mathcal{D} = U\, P_+$ and it commutes with the Hamiltonian. Furthermore, $\mathcal{D}$ flows to a non-invertible topological line operator of the Ising CFT (which has a non-trivial kernel in the space of conformal blocks) in the continuum limit. This lattice operator $\mathcal{D}$ is one of the most prominent examples of a non-invertible symmetry discussed in the literature.

The generalized symmetries of our model can be defined as the set of operators which
\begin{itemize}
    \item commute with the Hamiltonian,
    \item each of them can be written as a product of a unitary operator $U$ and a projector $P$ (so that they are physical operations which commute with time evolution), 
    \item are Hermitian so that they give the quantum numbers distinguishing all energy eigenstates of the theory, and
    \item {we include non-unitary operators in the set only if such operators are necessary for distinguishing different excitations of the theory.}
\end{itemize}
These operators form the symmetry monoid of our theory. Each element of the monoid can be interpreted as a unitary transformation before or after postselection conditioned by a suitable subspace (see \cite{Okada:2024qmk} for a discussion on how non-invertible symmetries of such types can be interpreted as quantum operations). {The third criterion above is applicable for the special case of stabilizer code models where the symmetry operations can also be used as measurements for detecting any excitation as discussed above in the context of the toric code (it is not a feature of general lattice models).

The last criterion above is needed to exclude projectors from the set of generalized symmetries in the (original) toric code where the usual elementary unitary (and Hermitian) stabilizers are sufficient to detect and distinguish all excitations of the theory. We will explicitly see that the stabilizer monoid (more below), which will be the subset of generalized symmetries that preserves all ground states and (as will be shown in Sec. \ref{Sec:Detect}) can \textit{distinguish various local excitations of our theory}, will necessarily include non-unitary (and Hermitian) operators. }

To see that the elementary plaquette operator $B_f$ is an operator of the type above, it is sufficient to note that 
\begin{equation}
    M_2 = U\, P, \,\, {\rm with}\,\, U = 1 \oplus Z, \,\, P = 0 \oplus I_{2\times 2},
\end{equation}
where $I_{2\times 2}$ is the $2\times 2$ identity operator.

\paragraph{\textbf{Stabilizer monoid}:} The stabilizer group of the toric code generalizes to the stabilizer monoid in our model. The stabilizer monoid $S_M$ is the set of operators satisfying the four properties mentioned above and furthermore which leave any ground state invariant. 

Clearly, the elementary stabilizers $\{A_v\}$ and $\{B_f\}$ are a subset of the generators of the stabilizer monoid $S_M$. Since, $A_v P_v = P_v$ (as $A_v^2 =1)$  and $B_fP_f = P_f$ (recall $B_f^3 =B_f$ and $B_f^4 = B_f^2$)
we obtain that
\be
A_v \mathcal{P}_0 = B_f \mathcal{P}_0 = \mathcal{P}_0,
\ee
where $\mathcal{P}_0$ is the projector to the ground state manifold, cf. Eq. \ref{Eq:P0gs}. The above implies that any ground state is invariant under the action of any $A_v$ and any $B_f$ since $\mathcal{P}_0\ket{G} =\ket{G}$ for any ground state $\ket{G}$. 

We can readily form the full stabilizer monoid $S_M$ which includes all symmetries that leave any ground state invariant by including more generators which are namely,
\begin{itemize}
    \item $M_{2,e}^2$ acting on any single edge $e$ of the lattice (note that $M_2^2$ commutes with $M_1$ so $M_{2,e}^2$ should commute with the Hamiltonian), and
    \item loop operators $C_{L_c, M_2}$ denoting $M_2$ acting on each edge of a contractible loop $L_c$ (note that any such loop operator can overlap with a vertex operator non-trivially on a pair of edges, and will therefore commute with any vertex operator and thus with the Hamiltonian). 
\end{itemize}

It is easy to see that since $M_2^2 = 0 \oplus I_{2\times 2}$ and it commutes with the vertex operators, we have $M_{2,e}^2 \mathcal{P}_0 = \mathcal{P}_0$, implying that it is a generalized symmetry.

In the toric code, it is not necessary to include $C_{L_c, M_2}$ as an additional generator of the stabilizer group for any contractible loop $L_c$ larger than the elementary plaquettes because any such loop operator is the product of the elementary plaquette operators enclosed by this loop. In our model, if we take such a product of all $B_f$ enclosed by $L_c$, then we get $C_{L_c, M_2}$ at the boundary times  products of $M_{2,e}^2$ on all edges enclosed by the loop. Nevertheless, the elementary plaquette operators $B_f$ are not invertible and therefore $C_{L_c, M_2}$ cannot be written in terms of the other generators. It follows that $C_{L_c, M_2}$ should be included as an additional generator of the stabilizer monoid.


We note that the stabilizer monoid $S_M$ is a commutative monoid since all the generators of the monoid commute with each other. We can also check that all elements of $S_M$ are generalized symmetry transformations. Furthermore, $S_M$ is not generated solely by the local operators, namely $\{A_v\}$ and $\{B_f\}$, as in the toric code, but also by non-local operators, namely $\{C_{L_c, M_2}\}$ of arbitrary sizes. However, it is also easy to see that, when restricted to the subspace spanned by the ground states (and also the fully mobile excitations discussed in the next section), $S_M$ is actually the same Abelian stabilizer group as in the case of the toric code, and is generated by the same elementary vertex and plaquette operators of the latter (composed of $X$ and $Z$ acting on each edge of the vertex or plaquette, respectively). 

We will explicitly show in Section \ref{Sec:Detect} that the stabilizer monoid can distinguish all the locally excited states of our model just like the stabilizer group in case of the toric code.

\begin{figure}[H]
\centering 
\includegraphics[scale=0.56]{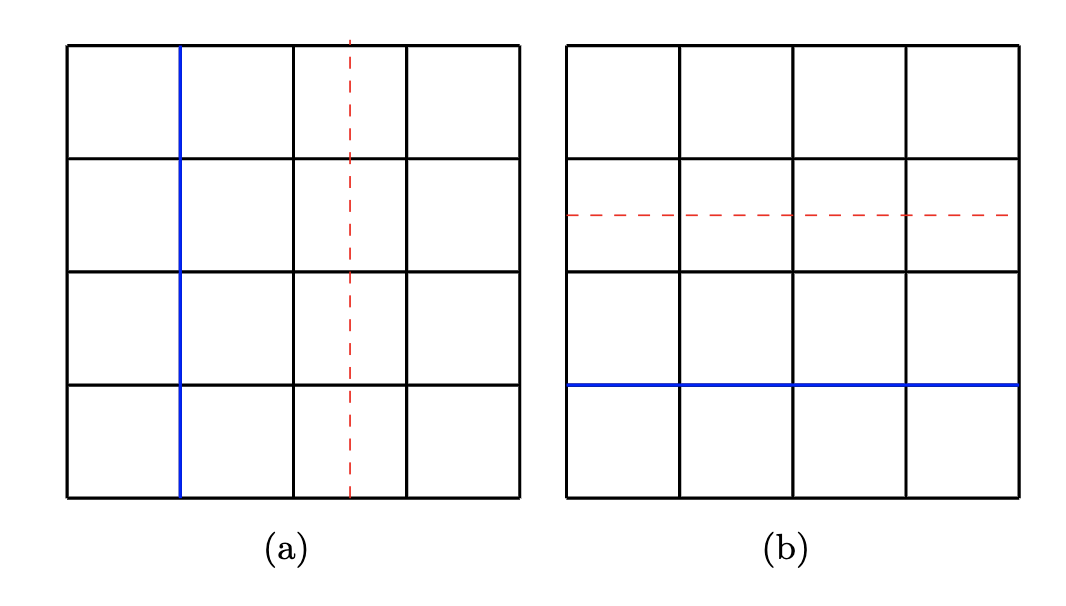}
\caption{(a) A non-contractible dual $M_1$ loop operator along the $A$ cycle of the torus $\tilde{C}_{A}$ is shown in dotted red, where we act with $M_1$ on every edge that is cut by the dotted line. A non-contractible direct lattice $M_2$ loop operator along the $A$ cycle is shown in blue. (Note that periodic boundary conditions have been imposed on the square lattice.) (b) A non-contractible dual $M_1$ loop operator along the $B$ cycle of the torus, $\tilde{C}_{B}$ is shown in dotted red. A non-contractible direct lattice $M_2$ loop operator along the $B$ cycle is shown in blue.}
\label{Fig:logical1}
\end{figure}

\paragraph{\textbf{Logical operators}:} Logical operators are generalized symmetries which have non-trivial actions on the ground states without annihilating them. These are operators $O$ that commute with all $A_v$ and $B_f$ (and thus with the Hamiltonian and $\mathcal{P}_0$), but $O \mathcal{P}_0 \neq \mathcal{P}_0$ and $O \mathcal{P}_0 \neq 0$. The set of these operators are generated by non-contractible loop operators $C_{L_{nc}}$ denoting the product of $M_2$ acting on each edge of a non-contractible loop $L_{nc}$ on the lattice, and $C_{\tilde{L}_{nc}}$ denoting the product of $M_1$ acting on each edge of a non-contractible loop $\tilde{L}_{nc}$ in the dual lattice. The total number of generators are thus $4g$. See Fig. \ref{Fig:logical1} for an illustration of the logical operators on a torus. These logical operators give a simple recipe for obtaining the $2^{2g}$ groundstates. Let us choose the seed state $\ket{S}$ to be 
\be\label{Eq:StateS}
\ket{S} = \bigotimes \prod_e \begin{pmatrix}
    0\\ 1 \\0
\end{pmatrix}_e
\ee
which is the $+1$ eigenstate of any $M_{2,e}$ and any $B_f$. Therefore, one ground state is
\begin{equation}
    \ket{G_1} = \mathcal{P}_0\ket{S} = \prod_v P_v \ket{S}.
\end{equation}
On the torus ($g=1$), as for instance, there are two independent non-contractible dual loops, placed on a $A$ cycle and a $B$ cycle, respectively, and therefore two independent $C_{\tilde{L}_{nc}}$, which we denote as $\tilde{C}_A$ and $\tilde{C}_B$ that cannot be transformed to each other via multiplication of $A_v$ operators. (The latter does not affect any groundstate as we have noted already.) Therefore, we get three other ground states which are 
\begin{align}
&\ket{G_2} =\prod_v P_v \tilde{C}_A \ket{S} = \tilde{C}_A \mathcal{P}_0 \ket{S}, \quad \ket{G_3} =\prod_v P_v \tilde{C}_B \ket{S} = \tilde{C}_B \mathcal{P}_0 \ket{S}, \nonumber\\ &{\rm and}\,\ket{G_4} =\prod_v P_v \tilde{C}_A \tilde{C}_B \ket{S} = \tilde{C}_A \tilde{C}_B \mathcal{P}_0 \ket{S}.
\end{align}
More generally, we will obtain $2^{2g}$ groundstates obtained from the $2^{2g}$ possible products of $2g$ non-contractible dual loop operators (including the identity) acting on the \textit{canonical} groundstate $\mathcal{P}_0\ket{S}$.\footnote{In the general case, we cannot consider a square lattice as it cannot tile a sphere or a compact Riemann surface with genus greater than $1$. Our model nevertheless can be defined on an arbitrary tiling of any Riemann surface.} Note that $C_{L_{nc}}$ keeps our chosen seed state invariant and cannot be used to generate any other groundstate. However, with a different choice of seed state we would have needed a different set of $2g$ logical operators for generating all the groundstates. 



In fact, when the action of the logical operators is restricted to the subspace spanned by the ground states, they reduce to the logical operators of the toric code; $C_{L_{nc}}$ reduces to the products of $Z$s acting on each edge of the non-contractible loop $L_{nc}$ while $C_{\tilde{L}_{nc}}$ reduces to the products of $X$s acting on each edge of the dual non-contractible loop $\tilde{L}_{nc}$. Therefore, on the subspace generated by the ground states, the logical operators can be identified with the normal subgroup formed by the quotient of the full symmetry group by the stabilizer group.

We will see that when restricted to the full sector of states that can decay to a ground state in presence of local perturbations (and composed of mobile electric and magentic charges), the symmetries, which form the stabilizer monoid and the set of logical operators, act as unitary transformations (representing the action of $\mathbb{Z}_2$). However, outside of this sector, the stabilizer monoid and the logical operators are not necessarily invertible.

\paragraph{\textbf{Annihilators}:} Interestingly, $K_e$ which denotes the action of $K = {\rm diag}(1,0,0)$ on an edge $e$ annihilates any ground state as well as the full sector of states that can decay to a ground state in presence of local perturbations. Since $K \,M_2 =0$, it follows that $K_e \mathcal{P}_0 =0$. $K_e$ commutes with both $M_1$ and $M_2$, and thus with the Hamiltonian. {One can actually define an annihilator monoid $S_M'$, which is the set of all Hermitian operators which commute with the Hamiltonian, and annihilate any ground state and also the full sector of fully mobile excitations. This monoid is generated by $\{K_e\}$ and non-vanishing products of any $K_e$ with any other generator of the symmetries which do not annihilate the ground state.} We will show that excited states with fractonic defects can also be eigenstates of $K_e$ with vanishing/non-vanishing eigenvalues depending on the edge $e$ where it is supported. {However, since the annihilators are not needed to detect various excitations of the theory, and because they are non-unitary operators, we should not include them in the list of generalized symmetries according to the last of the four defining criteria of generalized symmetries stated before.}  

\paragraph{\textbf{Summary}:} {The generalized symmetry monoid of our model is generated by the generators of the commutative stabilizer monoid which preserve all ground states, and the $2g$ logical operators, namely the non-contractible loops and dual loop operators which generates other ground states from a ground state. These symmetries act as unitary operators in the sub-space spanned by ground states and fully mobile excitations, but more generally these are non-invertible operators. Additionally, our model has Hermitian operators  $\{K_e\}$ which commute with the Hamiltonian, and annihilate all ground states and the sector of states with fully mobile excitations.}

\section{The excited states}\label{Sec:Excitations}

\subsection{Deconfined excitations}
\begin{figure}
    \centering
    \includegraphics[scale=0.56]{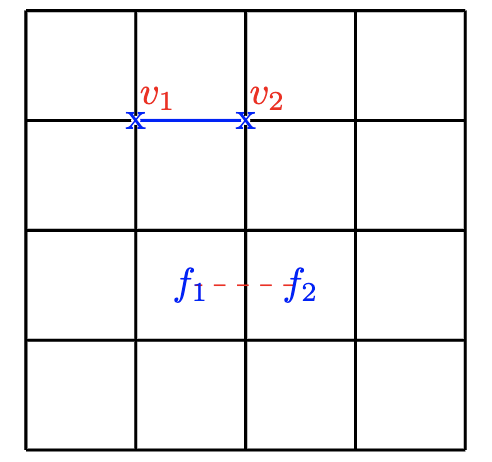}
    \caption{The smallest $M_1$ string (dashed red) excites two face operators at the faces $f_{1,2}$. The smallest $M_2$ string (blue) excites two vertex operators at the vertices $v_{1,2}$.}
    \label{Fig:exc1}
\end{figure}

The deconfined excitations of this model are similar to that of the toric code. Consider a string operator, $E_{v_1,v_2}$, which denotes the action of $M_2$ on each edge of a connected path on the lattice starting at the vertex $v_1$ and ending at the vertex $v_2$. See Fig. \ref{Fig:exc1} for an illustration of such a string operator of the smallest length, i.e. supported on a single edge. This string commutes with all vertex operators intersected by this path except $A_{v_1}$ and $A_{v_2}$. It anti-commutes with both $A_{v_1}$ and $A_{v_2}$, and therefore 
\be
E_{v_1,v_2} P_{v} = \begin{cases}
P^{\perp}_{v} E_{v_1, v_2},\, {\rm if} \,\, v = v_1 \,\,{ \rm or}\,\, v = v_2, \nonumber\\
P_{v} E_{v_1,v_2},\, {\rm otherwise, }\nonumber\\
\end{cases}
\ee
where $P^{\perp}_v = (1/2)(1- A_v)$ is the projector to the $-1$ eigensector of $A_v$. Consequently, for a generic seed state $\ket{s}$,
\begin{align}
    &\ket{v_1,v_2} = E_{v_1,v_2}\mathcal{P}_0 \ket{s} =  \mathcal{P}_{v_1,v_2} E_{v_1,v_2} \ket{s}, {\rm \,\,with }\nonumber\\
    & \mathcal{P}_{v_1,v_2} = P^\perp_{v_1} P^\perp_{v_2} \prod_{v \neq v_1, v_2} P_v \prod_f P_f 
    \label{Eq:exc1}
\end{align}
It is easy to see that $\mathcal{P}_{v_1,v_2}$ is the projector to the subspace with energy $+4$ above the groundstate in which two vertices, namely $v_1$ and $v_2$ are \textit{excited}. $\ket{v_1,v_2}$ is thus an energy eigenstate.

We can excite vertices only in pairs via action of such string operators $E_{v_1,v_2}$ on a ground state (which as discussed before always takes the form $\mathcal{P}_0 \ket{s}$). It can be explicitly checked that the rank of the projector which projects to states with an odd number of excited vertices is zero. As for instance,
\be
{\rm Tr}(\mathcal{P}_{v_1}) = 0, \,\, {\rm where}\,\,  \mathcal{P}_{v_1}=  P^\perp_{v_1}  \prod_{v \neq v_1} P_v \prod_f P_f 
\ee
Furthermore, the energy eigenvalue of $\ket{v_1,v_2}$ does not depend on the length of the string operator $E_{v_1, v_2}$ which creates it by acting on the ground state $\mathcal{P}_0 \ket{s}$. Thus such vertex excitations are deconfined. Following the terminology used in the context of the toric code, we will call these \textit{electric} excitations which always exist in pairs in this sector of purely deconfined excitations.

The other deconfined excitations are created by the dual string operators $\tilde{E}_{f_1,f_2}$ which denotes the action of $M_1$ on each edge of the real lattice intersected by a connected path in the dual lattice starting at face $f_1$ and ending at face $f_2$. See Fig. \ref{Fig:exc1} for an illustration of such a dual string operator of the smallest length, i.e. supported on a single edge. It is easy to see that it anti-commutes with only two plaquette operators, namely $B_{f_1}$ and  $B_{f_2}$, and commutes with any other plaquette operator. Therefore,
\be
\tilde{E}_{f_1,f_2} P_{f} = \begin{cases}
P^{\perp,-}_{f} \tilde{E}_{f_1, f_2},\, {\rm if} \,\, f = f_1 \,\,{ \rm or}\,\, f = f_2, \nonumber\\
P_{f} \tilde{E}_{f_1,f_2},\, {\rm otherwise, }\nonumber\\
\end{cases}
\ee
where $P^{\perp,-}_f = (1/2)(-B_f+ B_f^2)$ is the projector to the $-1$ eigensector of $B_f$. Consequently,
\begin{align}
    &\ket{f_1,f_2} = \tilde{E}_{f_1,f_2}\mathcal{P}_0 \ket{s} =  \mathcal{P}^-_{f_1,f_2} \tilde{E}_{f_1,f_2} \ket{s}, {\rm \,\,with }\nonumber\\
    & \mathcal{P}^-_{f_1,f_2} = \left(\prod_{v} P_v\right) P^{\perp,-}_{f_1} P^{\perp,-}_{f_2} \prod_{f\neq f_1, f_2} P_f 
    \label{Eq:exc2}
\end{align}
We note that $\mathcal{P}_{f_1,f_2}$ is a projector to the energy eigensector with energy $+4$ above the groundstate in which only the two faces, namely $f_1$ and $f_2$ are excited. $\ket{f_1,f_2}$ is thus an energy eigenstate. 

We can create such excited states with pairs of faces excited by the action of dual string operators on a ground state. Following the toric code terminology, these are the deconfined \textit{magnetic} excitations whose energy is independent of the length of the dual string creating them. 
The identities
\be
E_{v_1, v_2} E_{v_2,v_3} = E_{v_1,v_3}, \quad \tilde{E}_{f_1, f_2} \tilde{E}_{f_2,f_3} = \tilde{E}_{f_1,f_3}
\ee
are obvious. Furthermore, consider any pair of strings $E_{v_1, v_2}$ or a pair of dual strings $\tilde{E}_{f_1, f_2}$ with the same endpoints and such that the (dual) strings in the pair are homologous to each other. Their combined action is a contractible loop or dual loop which is part of the symmetry monoid $S_M$ that leaves any ground state invariant. Together these imply that we can move one or both of the endpoints of a string or a dual string operator via actions of operators of finite length and without any extra energy cost (meaning that by action of other (dual) string operators one can translate the end points of any (dual) string without creating more excitations). Thus the deconfined electric and magnetic excitations are mobile.

The whole sector of deconfined excitations (just like the ground states as mentioned before) is annihilated by the action of $K_e$ involving the action of the the matrix $K = {\rm diag}(1,0,0)$ acting on any edge $e$ of the lattice, since only the subspace spanned by $(0,1,0)$ and $(0,0,1)$ on each edge (the toric code subspace) appear in such states. The dynamics in this sector mimics the toric code exactly. For the same reasons, all elements of the stabilizer monoid and the logical operators act as unitary transformations representing the action of $\mathbb{Z}_2$ on this sector of states.

\subsection{Fractonic excitations: Confined immovable defects and their internal states}\label{Sec:Fracton}

This model admits another class of excitations that are immobile (fractonic) and confined. Note that for the deconfined excitations, the excited plaquettes have $-1$ eigenvalues for the respective face operators $B_f$.  Here we will study the excitations in which the vertices are not excited (i.e. all $A_v$ have $+1$ eigenvalues)  but in which some of the face operators $B_f$ have zero eigenvalues (zero fluxes), and show that such excitations are fractonic and confined. 

\begin{figure}
    \centering
    \includegraphics[scale=0.56]{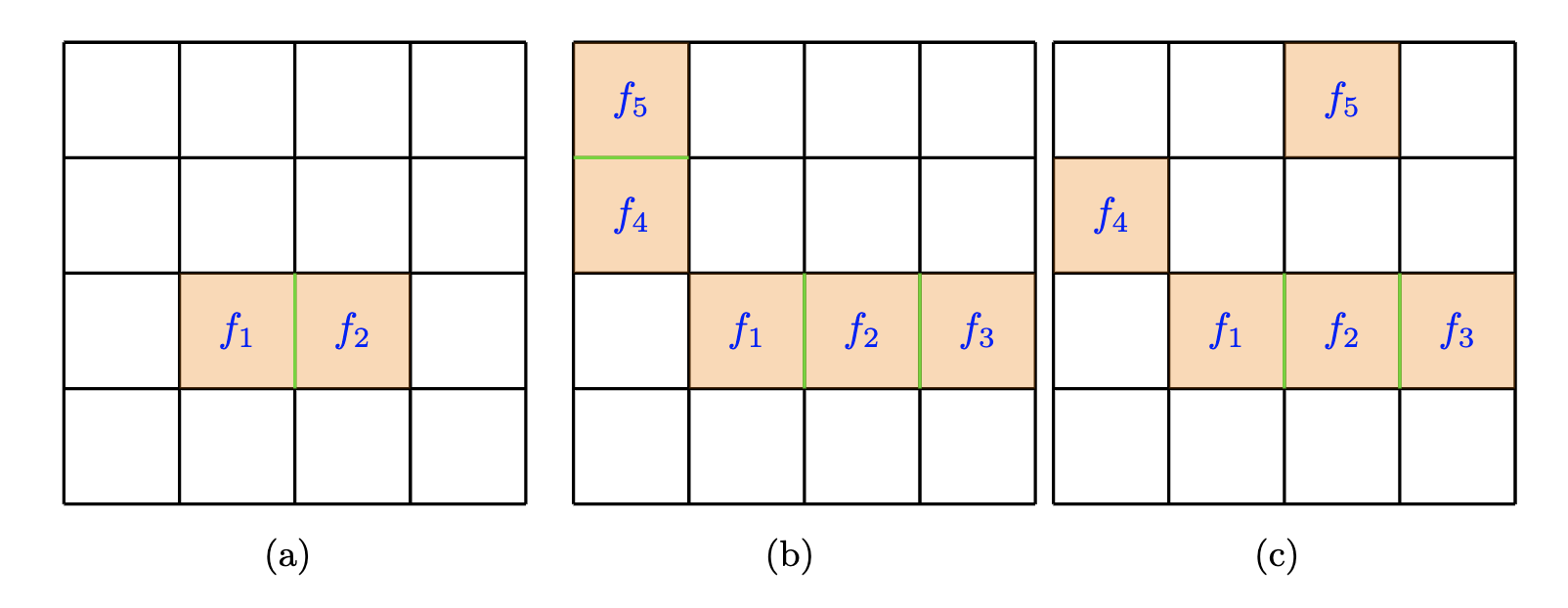}
    \caption{(a)Simplest allowed configuration of zero flux faces shown in orange. Green edge indicates the action of $M_3$.(b)Example of an allowed configuration of zero flux faces where all of them share an edge with at least one other zero flux face. (c) Example of a disallowed configuration where the zero flux faces $f_4$ and $f_5$ don't share an edge with any other zero flux face excitation.}
    \label{Fig:exc3}
\end{figure}

The projector to the zero eigenvalue sector of $B_f$ is $P^{\perp,0}_f = (1-B_f^2)$. Let us consider the following projector   $\mathcal{P}^{d}_{f_1,f_2, \cdots, f_n}$ to the energy eigenstate sector in which $n$ faces $f_1, f_2, \cdots, f_n$ are excited to zero eigenvalues:
\be
\mathcal{P}^{d}_{f_1,f_2, \cdots, f_n} = \left(\prod_{v} P_v\right) P^{\perp,0}_{f_1}P^{\perp,0}_{f_2}\cdots P^{\perp,0}_{f_n} \prod_{f\neq f_2,f_2, \cdots f_n} P_f.
\ee
As illustrated in Appendix \ref{App:Trace}, ${\rm Tr}(\mathcal{P}^{d}_{f_1,f_2, \cdots, f_n}) =0$ if any of the faces $f_i$ with zero flux does not share an edge with another face $f_j$ with zero flux ($i,j = 1, \cdots, n$ and $i\neq j$). Thus any isolated zero flux face excitation cannot exist. The trace (and thus the rank of the projector) is non-vanishing otherwise implying existence of eigenstates with energy $+n$ above the ground state. See Fig. \ref{Fig:exc3} for an illustration.
Which operator creates such excitations? Let us consider the simplest excitation of this type in which two adjacent faces $f_1$ and $f_2$ have zero fluxes (see Fig. \ref{Fig:exc3} (a)). The rank of the projector to this energy eigenstate sector is $2^{2g}$ (see Appendix \ref{App:Trace}). Therefore, it is natural to postulate that these energy eigenstates can be created by an operator $M_{3}$ supported on the edge $e'$ shared by $f_1$ and $f_2$ acting on any of the $2^{2g}$ groundstates.

To construct $M_3$ explicitly, it is convenient to consider the seed state $\ket{S}$ defined in Eq. \eqref{Eq:StateS} which is a $+1$ eigenstate of all $B_f$. Demanding that $M_3$ commutes with $M_1$ and that it rotates the $+ 1$ eigenstate of $M_2$ to the $0$ eigenstate of $M_2$, we obtain (in the Pauli $Z$ basis)
 \begin{equation}\label{Eq:M3}
     M_3 = 
    \begin{pmatrix}
        0&1&-1\\
        0&0&0\\
        0&0&0
    \end{pmatrix}.
    \end{equation}
Note $M_3$ endowed with the properties that we have demanded is not unique, however we will discuss the rationale for the above choice in the following section. For the present purposes, any other choice of $M_3$ would also work. Let us denote the operator $M_3$ supported on an edge $e'$ as $M_{3,e'}$. It follows that with $\ket{S}$ given by Eq. \eqref{Eq:StateS}, we obtain 
\begin{align}
& \ket{E} = M_{3,e'}\mathcal{P}_0\ket{S} = M_{3,e'}\prod_v P_v \left(\bigotimes \prod_e \begin{pmatrix}
    0\\ 1 \\0 \end{pmatrix}_e\right) = \prod_v P_v M_{3,e'}\left(\bigotimes \prod_e \begin{pmatrix}
    0\\ 1 \\0 \end{pmatrix}_e\right),\nonumber\\
&= \prod_v P_v \ket{S'}, \,\, {\rm with} \,\,\ket{S'} =\left(\begin{pmatrix}
    1\\ 0 \\0\end{pmatrix}_{e'} \bigotimes\prod_{e\neq e'} \begin{pmatrix}
    0\\ 1 \\0\end{pmatrix}_e\right).
\end{align}
The second equality above follows from the fact that $\ket{S}$ is left invariant by any $P_f$. The third and fourth equalities follow from the defining properties of $M_3$ that it commutes with $M_1$, and it takes the $+1$ eigenstate of $M_2$ to the $0$ eigenstate of $M_2$, respectively (thus any such $M_3$ which can be different from the choice in Eq. \ref{Eq:M3} will also work). We can show that $\ket{E}$ is an eigenstate of the Hamiltonian. The final form of $\ket{E}$ implies that it is a $+1$ eigenstate of all $A_v$ since the product of all $P_v$ acts on the extreme left. Furthermore, as $B_f$ commutes with all $A_v$, we obtain
\begin{equation}
    B_f \ket{E} \begin{cases} = \ket{E}\,\, {\rm if \,\,  the\,\,  face}\,\,  $f$ \,\,  {\rm does\,\,  not\,\,  have\,\,  the\,\,  edge\,\,}$e'$,\\
    =0, \,\, {\rm otherwise}\end{cases}.
\end{equation}
Thus $\ket{E}$ is an energy eigenstate in which all $A_v$ have $+1$ eigenvalues, and all faces $B_f$ have $+1$ eigenvalues except for the two faces $f_1$ and $f_2$ shared by the edge $e'$ (on which the action of $M_3$ has been supported). The other independent states in the image of $\mathcal{P}^d_{f_1, f_2}$ can be created by the action of $M_{3,e'}O$ (or $O M_{3,e'}$) acting on $\mathcal{P}_0\ket{S}$ where $O$ is a logical operator which is made out of products of $M_1$ acting on each edge of a non-contractible loop on the dual lattice, or a product of such logical operators. We can thus account for the $2^{2g}$ degeneracy of the smallest element of the sector with two adjacent zero flux faces.

As evident from the result of the explicit computation of the rank of the projector $\mathcal{P}^d_{f_1,\cdots, f_n}$, the excitation pair created via the action of $M_3$ on a single edge is confined. As for instance, consider the pair of adjacent zero flux faces $f_4$ and $f_5$ in Fig. \ref{Fig:exc3}(b). This pair cannot be separated without creating additional zero flux excitations since two separated face excitations with zero fluxes are disallowed. The separation of a pair of adjacent zero flux excitations can be achieved only by a creation of a string of zero flux excitations between them costing energy proportional to the length of the string. This is the hallmark of confinement.\footnote{Recently, it has been pointed out that an Hamiltonian with fractonic excitations can be obtained by integrating out a confining gauge field \cite{Pai:2019hor}. A similar phenomena also occurs for the fluxes in a $\mathbb{Z}_2$-toric code with matter fields on the vertices as discussed in \cite{ferreira2015recipe}.} 

Since a pair of zero flux faces cannot be separated from each other via local operations without encountering energy cost, these are fractons. Larger clusters of zero flux excitations (more about them below) are also fractons because such clusters cannot be fragmented into separated parts via local operations without encountering energy cost. {We will give a precise definition of \textit{fractonic excitations} in terms of their immobility in Sec. \ref{Sec:Fusion} where we will show that the fractonic excitations cannot be moved by the operators in the \textit{spectral monoid} which generates all local excitations of the theory by acting on the ground state. In this sense, the fractonic excitations of our model are completely immobile, i.e. they cannot be moved even accounting for energy cost. Furthermore, in Sec. \ref{Sec:Fusion}, we have also discussed how the fusion rules can be used more generally to define the fractonic nature of excitations.}

{Another remarkable property of the fractonic confined excitations of our model is that these have internal degrees of freedom when they are constituted out of a sufficient number of zero flux faces forming clusters with specific geometric shapes.}

\begin{figure}
    \centering
    \includegraphics[scale=0.56]{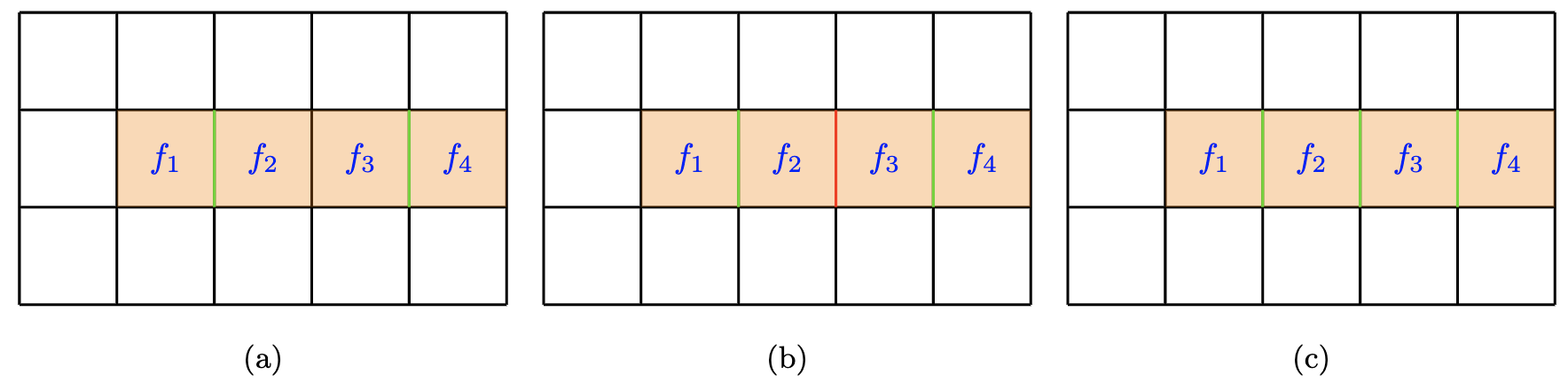}
    \caption{Three states with the faces $f_{1,2,3,4}$ excited to the zero eigenvector. Green lines indicate the action of $M_3$ on the edge and red lines denote the action of $M_1$.}
     \label{Fig:exc4a}
\end{figure}

Consider three zero flux excitations $f_1$, $f_2$ and $f_3$ forming a linear chain as shown in Fig. \ref{Fig:exc3a}.  The trace of the corresponding projector $\mathcal{P}^d_{f_1,f_2,f_3}$ is $2^{2g}$ (see Appendix \ref{App:Trace}) implying that there is no internal degree of freedom for this zero flux cluster. These states are created via the product of $M_3$ supported on the two internal edges  (i.e. that shared between $f_1$ and $f_2$ and that shared between $f_2$ and $f_3$, respectively) acting on one of the $2^{2g}$ groundstates. The conclusion is the same if these three adjacent zero flux states is L-shaped instead of being a linear chain.
\begin{figure}
    \centering
    \includegraphics[scale=0.56]{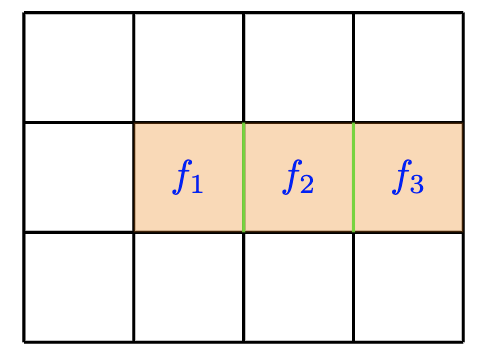}
    \caption{Three faces $f_{1,2,3}$ are excited to the zero eigenstate.}
 \label{Fig:exc3a}
\end{figure}
However, the case with four zero flux face excitations is less trivial. Consider four face excitations $f_1$, $f_2$, $f_3$ and $f_4$ forming a linear chain as shown in Fig. \ref{Fig:exc4a}. The rank of the corresponding projector $\mathcal{P}^d_{f_1,f_2,f_3,f_4}$ is $3 \times 2^{2g}$ as shown in Appendix \ref{App:Trace}. The factor of $3$ can be accounted for by $3$ distinct operators as shown in Fig. \ref{Fig:exc4a}.
Firstly, after the action of $M_1$ on an edge shared by two zero flux faces, the latter remain zero flux faces. Since $M_3 M_1 = -M_3 = M_3 M_1$, therefore $M_1$ acting on an edge where the action of $M_3$ is supported cannot create a new state. {Furthermore, acting with $M_1$ on any other internal edge of a zero flux face cannot change the eigenvalue (equal to zero) of the corresponding $B_f$ operator since one edge of the face, where $M_3$ acts, is already in the zero eigenstate of $M_2$. However, a new internal state of the defect may be generated by such an action.\footnote{Note that $M_1$ cannot act on any of the boundary edges of the orange defect region since that will excite the adjoining face outside the region to the $-1$ eigenvalue. This actually creates magnetic monopoles as discussed in detail in the following subsection.} The factor of $3$ in the degeneracy arises from the $3$ possible actions of $M_3$ and $M_1$ as shown in Fig. \ref{Fig:exc4a}. (The central internal edge can have identity, $M_1$ or $M_3$ acting on it while both the other two internal edges must support the action of $M_3$ so that four zero flux face excitations are created as shown in the figure.) Remarkably, a linear chain of four zero flux face excitations (with no vertex excitations) thus effectively acquires a spin $1$ internal degree of freedom. 

We can also construct a configuration of four zero flux faces forming a square as shown in Fig. \ref{Fig:exc4b}. As shown in Appendix \ref{App:Trace}, the rank of the corresponding projector $\mathcal{P}^d_{f_1,f_2,f_3,f_4}$ is $9\times 2^{2g}$ implying that this zero flux configuration has $9$ internal states (and thus effectively a spin $4$ internal degree of freedom). 

\begin{figure}[H]
    \centering
    \includegraphics[scale=0.86]{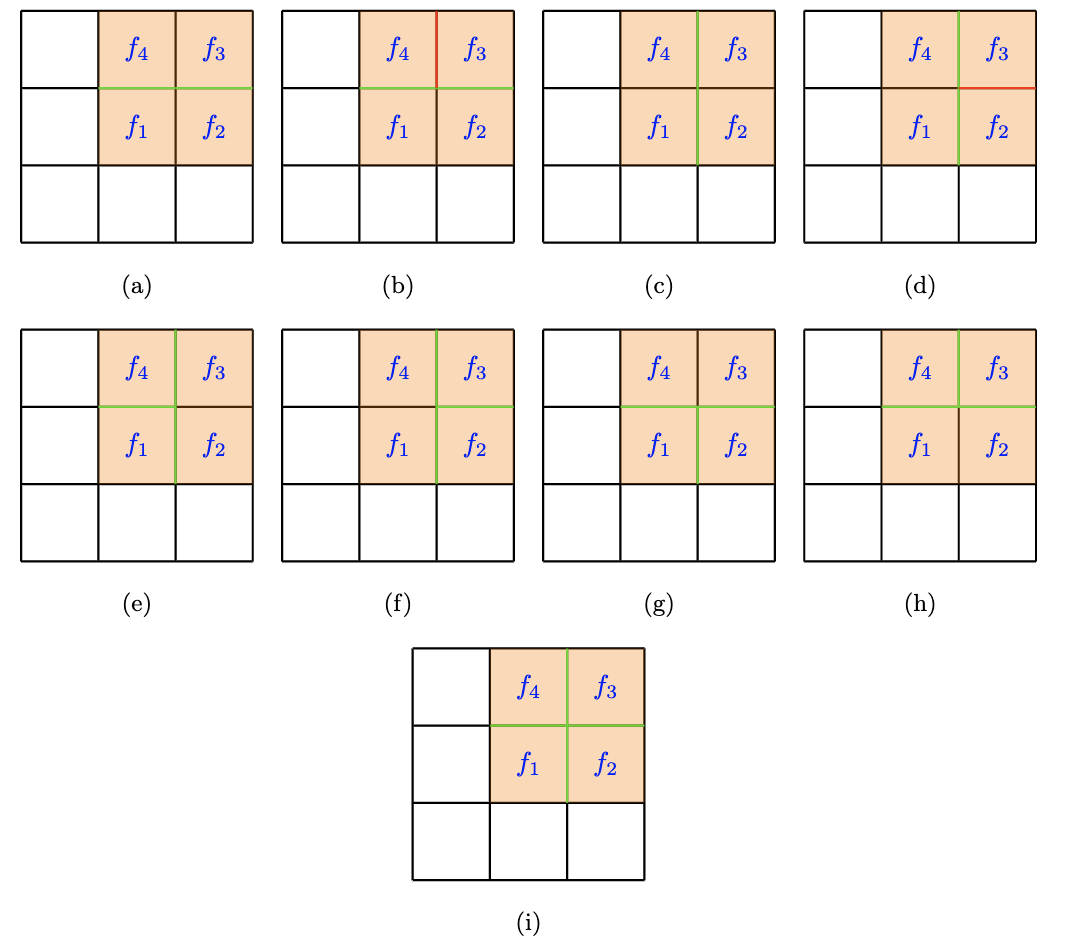}
    \caption{Nine distinct configurations with four face excitations arranged in a square. Solid green lines indicate the action of $M_3$ and red lines indicate the action of $M_1$.}
    \label{Fig:exc4b}
\end{figure}
The $9$ operators which can act on one of the $2^{2g}$ groundstates to create these linearly independent $9 \times 2^{2g}$ excited states involving inequivalent combinations of $M_3$ and $M_1$ are shown in  Fig. \ref{Fig:exc4b}. Since $A_v$ leaves any ground state invariant, if two operators $O$ and $O'$ are related by $O = O' A_v$, then they will create the same state acting on any groundstate and will be equivalent (for a precise definition of equivalence see the following section). Using $M_1^2 =1$ and $M_1 M_3 = -M_3 = M_3  M_1$, we thus obtain only two inequivalent operators involving a product of the action of $M_3$ on two internal edges and that of $M_1$ on another internal edge. Similarly, the action of $M_1$ on an internal edge does not produce any new state if $M_3$ acts on the remaining three internal edges. We thus explicitly obtain the $9$ operators which together with the logical operators generate all the excited states in which four zero flux face excitations form a square and in which no vertex is excited.

Unlike other stabilizer code models, fractonic zero flux face configurations (defects) do not acquire partial or complete mobility when they combine to form bigger clusters. {This implies that our fractonic model is not of type I. {(See next subsection to see why our model is also not of type II.) }However, as demonstrated above, fractonic defects acquire internal degrees of freedom where the number of internal states depends on the geometry and size of the configuration.} These internal states are generated by the different combinations of $M_3$ and $M_1$ that can be supported on the internal edges (always accompanied by the actions of $M_3$ supported on appropriate number of internal edges creating the zero flux configuration) acting on any vacuum state.

{It is not clear if the fractonic defects of our model can be associated with dipole conservation or similar conserved charges of sub-system symmetries. Such symmetries impose kinematic constraints on fractonic excitations but do not make them completely immobile (note dipole charge conservation does not restrict movement of the center of mass of a dipole provided its orientation is unchanged) \cite{Nandkishore:2018sel,Pretko:2020cko}. The fractonic clusters of arbitrary shapes and sizes in our model are completely immobile but their internal degeneracy aside from the topological part (as discussed in the above examples) is not related to kinematic constraints but rather to different possibilities of disentangling the local degrees of freedom from a parent ground state.} 

$K_e$ involving the action of the the matrix $K = {\rm diag}(1,0,0)$ supported on any edge $e$ of the lattice are symmetries have interesting actions on the states featuring fractonic defects. If it acts on any edge except for the edges supporting the action of $M_3$ needed to create the fractonic defect configuration, the state is annihiliated. Otherwise the state is an eigenstate of this symmetry. Therefore, the action of $K_e$ can detect not only the location of the fractonic defects, but also partial information of how it is created via the action of $M_3$ supported on the internal edges of the configuration. Similarly, $M_2^2$ acting on a single edge will annihilate any defect configuration if it overlaps with any edge where $M_3$ needs to act to create the defect configuration from a ground state. We will show in Sec. \ref{Sec:Detect} that the internal states of any fractonic cluster can be detected and distinguished by measurements of the generators of the commutative stabilizer monoid.

\subsection{Deconfined stuff and fractons: Magnetic monopoles and restricted electric charges}

Here we analyze the general case where we can have vertex excitations, i.e. vertices $v$ with $A_v$ eigenvalues $-1$ together with face excitations of both types, i.e. faces with $B_f$ eigenvalues which can be either $-1$ or $0$. Such excited states involve both electric and magnetic charge excitations and zero flux configurations (defects). Since we have already studied which operators can create such excitations by acting on a vacuum state, we can readily deduce how such general excitations can be created. Remarkably, we find that the mobile excitations, i.e. the electric and magnetic charges change their nature in the presence of the fractonic defect configurations. 

Firstly, a fractonic defect configuration can absorb any of the magnetic charges of a pair (created by a dual string) and thus create magnetic monopoles. Consider the configuration shown in Fig. \ref{Fig:monopole}. Here $M_1$ acts on the outer edge of a fractonic defect configuration. This is the shortest dual string connecting two adjacent faces, but with one of the faces in the defect region. Here the action of $M_1$ excites only the face exterior to the defect region to $-1$ eigenvalue of $B_f$. As discussed before, the $B_f$ eigenvalue of the other face within the defect region remains zero. This implies that we have created a magnetic monopole attached to the defect region. 

\begin{figure}[H]
    \centering
    \includegraphics[scale=0.56]{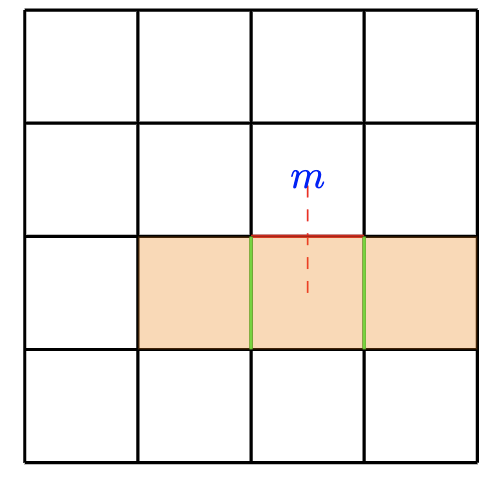}
    \caption{The defect region (orange) absorbs one magnetic charge creating an isolated magnetic charge $m$. Note that the other face connected to the edge with the red $M_1$ operator is not excited to the $-1$ eigenvector since that face is already in the defect region and excited to the $0$ eigenvalue.}
 \label{Fig:monopole}
\end{figure}
More generally, we can consider any dual string $\tilde{E}_{f_1, f_2}$ with one of the terminal faces, as for instance $f_2$, coinciding with a face at the border of a defect region with zero flux. It is easy to see that in such a case only the face $f_1$ will be excited to $-1$ eigenvalue of $B_f$, whereas the face $f_2$ will still have zero flux. In the presence of a zero flux defect configuration, we can thus create a magnetic monopole (where a single face is excited to $-1$ eigenvalue) anywhere outside the defect configuration. 

Furthermore, if the dual string $\tilde{E}_{f_1, f_2}$ has one of the terminal faces, as for instance $f_2$, not at the border but within a defect region with zero flux, then via the action of $M_1$ on an internal edge, it may modify the internal state of the defect while creating a magnetic monopole (at the face $f_1$) outside the defect configuration. Therefore, dual string operators $\tilde{E}_{f_1, f_2}$ (magnetic charges) interact non-trivially with zero flux defect configurations since one of the endpoints of a dual string can be absorbed by the defect configuration with or without altering the internal state of the latter. 

{In fact, we can consider the fractonic defects as regions where the magnetic charges have condensed simply because the action of $M_1$ on any edge within the defect region either leaves the state unaltered or creates a new state with the same energy. This implies that the expectation value of $M_{1,e}$  (the shortest dual string operator associated with a pair of magnetic charges) and also a dual string operator localized in the defect region can be non-vanishing in the fractonic cluster state if the edges supporting these operators are appropriately chosen.\footnote{Again this follows from $M_1 M_3 = M_3 M_1 = - M_3$. Consider the fractonic state $\ket{f_e} = M_{3,e} \mathcal{P}_0 \ket{s}$ in the basis where $M_2$ is diagonal. We readily see that $\bra{f_e} M_{1,e} \ket{f_e} = -1$. However, $\bra{f_e} M_{1,e'} \ket{f_e} = 0$ if $e\neq e'$. } Furthermore, the expectation values of the dual (magnetic) strings localized within the defect region distinguishes the various internal states of the fractonic defect states with zero flux faces in the defect region. We will have more to say on this issue in Sec. \ref{Sec:Decon} where we discuss a phase transition in our model which is produced by adding a new term in the Hamiltonian.}

\begin{figure}[H]
    \centering
    \includegraphics[scale=0.56]{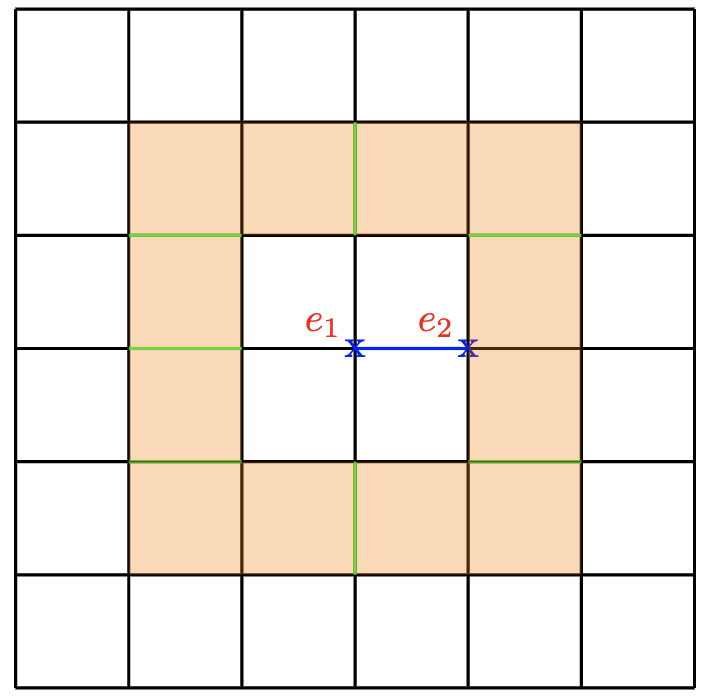}
\caption{The defect region (orange) contains a pair of electric charges $e_{1,2}$ created by the action of $M_2$ (blue) on an edge. These charges have restricted mobility and can only be moved along paths that don't include the green edges with the action of $M_3$.}
 \label{Fig:confcharge}
\end{figure}
To analyze the nature of electric charge excitations in presence of zero flux defect configurations, we need to be aware that $M_2$ and $M_3$ do not commute, and that $(M_2\, M_3)_e \ket{G} = M_{2,e}\ket{E} = 0$ since the action of $M_3$ supported on an edge $e$ on any ground state brings the state of the edge to the $0$ eigenstate of $M_2$ (see Eq. \ref{Eq:M3}). Let us consider operators creating energy eigenstates by acting on the ground state such that those creating zero flux configurations (the actions of $M_3$ and $M_1$ on various edges) are to the right of the string operators $E_{v_1, v_2}$ which create pairs of electric charge excitations. In this context, the zero flux configurations act as perfect repellers of electric charges. Consider the configuration shown in Fig. \ref{Fig:confcharge}. The charges, $e_1$ and $e_2$ at the ends of the string can only move along a path on the lattice which does not intersect on any edge where $M_3$ acts (as otherwise the state is annihilated). Thus electric charges acquire restricted mobility in presence of the zero flux configurations if they are created after the creation of the defect configuration (and the full operator creating the entire configuration is ordered in the way mentioned). \textit{Unlike the magnetic monopole, the electric charge cannot be passed through a defect region along an arbitrary path.} On the other hand, if we allow other operator orderings, then we create hybrid configurations in which vertices within defect configurations can be excited.

Although such configurations of magnetic monopoles and electric charges external to fractonic defects can retain mobility in their local neighborhoods but full configurations of such types form superselection sectors distinct from that of the ground state as will be shown in Sec. \ref{Sec:FRNI}. Therefore, our fractonic model is not of type II. Earlier we have argued that our model is not of type I either. We will also find that the excitations of our model are described by a novel fusion category which is both non-commutative and non-Abelian, and requires including the zero operator for closure.

\subsection{Detecting the excitations using the stabilizer monoid}\label{Sec:Detect}
The quantum error correction literature describes the detection of excited states within the stabilizer group formalism \cite{Gottesman:1997zz}. One important feature of stabilizer codes is that the stabilizer group can be utilized to distinguish all local excitations, i.e. excited states which are created by the action of operators supported on a bounded sub-region on a ground state. Measurements of the different elements of the Abelian stabilizer group can exactly characterize any local excitation. In our model, such a role is played by the commutative stabilizer monoid which are the generalized symmetries which leave any ground state invariant and which have been discussed in detail in  Sec. \ref{Sec:GenSym}.

We recall that the elementary stabilizers $A_v$ and $B_f$ are among the generators of the commutative stabilizer monoid. Measuring the operator $A_v$ yields two possible outcomes $\pm 1$, where the $-1$ outcome indicates the presence of an electric charge excitation at the vertex $v$. Similarly, the operator $B_f$ yields the measurement outcomes $\pm 1, 0$, where the $-1$ outcome indicates the presence of a (mobile) magnetic flux excitation at the face $f$ and the $0$ outcome indicates the presence of an immobile (fractonic) zero flux excitation. 

However, measurements of $A_v$ and $B_f$ cannot detect the internal excited states of fractonic clusters, eg. those shown in Fig. \ref{Fig:exc4a}. To distinguish the fractonic excitation explicitly, it is not enough to know which faces have zero flux, but also precisely which edges support the actions of $M_3$ and $M_1$ that are  responsible for the creation of the fractonic cluster state from a parent ground state. The other generators of the stabilizer monoid, namely $M_{2,e}^2$ (the action of $M_2^2$ on arbitrary edges) and $C_{L_c,M_2}$ (the action of $M_2$ on the edges of arbitrary contractible loops $L_c$ larger than the elementary plaquettes) discussed in Sec. \ref{Sec:GenSym} are sufficient for this purpose. The edges which support the actions of $M_3$ can be detected by measuring $M_{2,e}^2$. If $M_3$ acts on the edge, the corresponding eigenvalue of $M_{2,e}^2$ is $0$, and otherwise it is $1$.  Measurement of various $C_{L_c,M_2}$ can be utilized to detect the edges which support the action of $M_1$ eg. in configurations depicted in  Fig. \ref{Fig:exc4a}. Consider a big $M_2$ contractible loop which encircles the defect region shown in in  Fig. \ref{Fig:exc4a} such that it intersects the defect region only on the middle internal edge. Depending on whether $1$ or $M_1$ or $M_3$ acts on this edge to create the  internal state of this defect, the eigenvalue of the $M_2$ loop would be $1$, $-1$ or $0$. Similarly, the collection of all generators of the commutative stabilizer monoid can distinguish mixed excitations containing fractonic clusters and mobile excitations (eg. magnetic dipoles and monopoles) localized outside the fractonic clusters.

\subsection{Deconfining the fractons}\label{Sec:Decon}
An interesting question is whether we can drive our model through a critical point where the fractons can deconfine. Consider the following modification to the Hamiltonian of our model
\begin{equation}\label{Eq:NewH}
    H' = -\sum_v A_v -\sum_f B_f -\lambda \sum_f  P^{\perp,0}_f.
\end{equation}
We recall that $P^{\perp,0}_f =\left(1-B_f^2\right)$ is the projector to the zero eigenspace of $B_f$. Above $\lambda$ is a real coupling constant. This coupling leads to \textit{frustration} because $P^{\perp,0}_f P_f = P^{\perp,0}_f B_f = 0$. So, we cannot  achieve the eigenvalue $1$ for both $B_f$ and $P^{\perp,0}_f$ for a given face $f$. Nevertheless, it is easy to see that the modified Hamiltonian remains a sum of commuting Hermitian operators, and hence we can readily solve this exactly.

The ground states of the original Hamiltonian (where $\lambda =0$) remain eigenstates of the modified Hamiltonian with energy
\begin{equation}
    -|v| - |f|.
\end{equation}
Consider a fracton configuration created by acting $M_3$ on some edges. This will create $|f_0|$ zero flux faces, and the energy of such a configuration computed with the  Hamiltonian \eqref{Eq:NewH} is
\begin{equation}
    -|v| - (|f|-|f_0|) - \lambda |f_0| =  -|v| - |f| - (\lambda -1) |f_0| .
\end{equation}
Clearly, if $\lambda <1$, then the energy of a configuration with zero flux faces is always higher than that of any ground state of the original Hamiltonian. Therefore, the ground states with topological order for $\lambda =0$ continue to be ground states for non-vanishing $\lambda$ as long as $\lambda <1$. When $\lambda >1$, the creation of zero flux faces lower the energy. Therefore, for $\lambda >1$, all faces of the lattice have zero flux in any ground state. 

A ground state at $\lambda > 1$ can be created by the action of $M_3$ on an original ground state (at $\lambda =0$) supported on sufficient number of edges on the lattice such that any face of the lattice overlaps with such an edge. This ensures that all faces on the lattice have zero flux. Following our previous analysis of fractonic clusters appearing at $\lambda =0$, further actions of $M_3$ or $M_1$ on the complementary set of edges (i.e. edges where $M_3$ has not yet acted) do not affect the energy of the state and thus leads to further ground state degeneracy. It also follows that the dual strings $\tilde{E}_{f_1,f_2}$ create \textit{gapped} deconfined excitations (pairs of deconfined magnetic charge excitations) when $\lambda <1$, but create new orthogonal ground states or leave the ground state invariant when $\lambda >1$. The ground state expectation value of $M_{1,e}$ (the shortest dual string) or an arbitrary $\tilde{E}_{f_1,f_2}$ can therefore be non-trivial if the edges which support these operators are appropriately chosen. Thus the system is in a phase where magnetic charges have condensed for $\lambda > 1$ and with the dual string operators acting as order parameters distinguishing different ground states.

The case $\lambda =1$ is the most interesting. In this case, the phase is best described as a \textit{fracton liquid}. We can readily see that a state of the form
\begin{equation}
    \ket{G} = \prod_v P_v\prod_{f\in \overline{F}} P_f \prod_{f\in F}P^{\perp,0}_f \ket{S}
\end{equation}
is a ground state with $\ket{S}$ being any suitable seed state. Above $F$ is a subset of faces in the lattice and $\overline{F}$ is the complement of $F$. As discussed before, the set $F$ should not contain any isolated zero-flux face which does not share any edge with another zero flux face, as otherwise the rank of the projector $$\prod_v P_v\prod_{f\in \overline{F}} P_f \prod_{f\in F}P^{\perp,0}_f $$vanishes. So, $F$ should be composed of clusters of zero flux faces which have no disconnected components with a single face. 

We readily note that since the choice of $F$ is not unique, the projector to the full ground state manifold at $\lambda =1$ is the sum of projectors corresponding to distinct choices of $F$ (note that all such projectors are mutually orthogonal to each other). Furthermore, for each choice of $F$, we have internal degeneracy as discussed before as the rank of the projector for a given choice of $F$ is a non-trivial multiple of $2^{2g}$. Since the size of $F$ can be extended arbitrarily with no energy cost, the fractons deconfine at $\lambda = 1$. 

Within each fractonic cluster the magnetic charges condense. However, we note that at $\lambda =1$, we can create ground states which are superposition of fractonic clusters localised at different regions of the lattice (like a cat state). Therefore, a generic ground state at $\lambda= 1$ has no local order parameter.

We will present a more comprehensive analysis of different deformations of the model and associated phases in a later work. 

\section{A novel fusion category: Non-commutative and non-Abelian}
\label{Sec:Fusion}
\subsection{An intuitive discussion}\label{Sec:Intutitive}
Superselection sectors and their products given by the fusion rules are important tools for classifying quantum phases of matter in terms of their fundamental excitations. We will briefly review first how this works for the toric code in an intuitive way and then we will proceed to give formal definitions which generalize the standard constructions in a way that can include non-invertible operators. For this entire section, we will consider a square lattice with a boundary and our statements should be understood in the limiting situation where the lattice is infinite (has infinitely many edges). Our discussion in this subsection will be similar to that in \cite{Pai:2019fqg} (see also \cite{haah2013commuting,haah2016algebraic}). Particularly in \cite{Pai:2019fqg}, a natural language has been developed to include fractonic excitations in superselection sectors and their fusion products so that one can understand the phase in a comprehensive way. In the following subsection, we will develop a rigorous methodology to include the feature of our model that the fractonic excitations are created by non-invertible operators.

Consider a bounded region $R$ in the lattice. Any string operator $E_{v_1, v_2}$ or dual string operator $\tilde{E}_{f_1, f_2}$ acting on the ground state will create a pair of magnetic and a pair of electric charge excitations, respectively. If $v_1, v_2 \in R$ and $f_1, f_2\in R$, then the reduced density matrix of these excited states obtained by tracing out the edges in $R$, will be the same as that of the vacuum (as evident from the explicit expressions \eqref{Eq:exc1} and \eqref{Eq:exc2}). Such excitations are thus localized in $R$. Furthermore, these states can be brought back to the ground states by the application of another $E_{v_1, v_2}$ or another $\tilde{E}_{f_1, f_2}$ both of which are entirely supported within $R$ and are homologous to the original string or dual string operators that created these excited states, respectively (although the latter need not be supported entirely within $R$). Therefore, all such pairs of electric and magnetic charge excitations with the excited faces and vertices, respectively, localized in $R$, are in the ground state super-selection sector because,
\begin{itemize}
    \item these excitations can be brought back to a ground state by the action of operators supported within $R$, and
    \item the reduced density matrix in the complement of $R$ is the same as that of a ground state.
\end{itemize}
We emphasize that the above does not imply that these states themselves are created necessarily by operators that are supported within $R$ by their action on ground states. This superselection sector can be referred to be that of $1$, the identity operator, which leaves the ground states invariant (we will give a precise definition of equivalence below).

The toric code however is a topological phase where non-local operators of arbitrary length play a fundamental role. Therefore, we expect that other superselection sectors exist and this is indeed the case. Consider a long string or a long dual string with one endpoint at infinity and another in $R$. This will create a single electric or magnetic charge excitation within $R$ and another electric or magnetic charge excitation which is infinitely far away. It is easy to see that
\begin{itemize}
    \item for such excited states the reduced density matrix over any region $R'$, which does not overlap with $R$, is at finite distance from $R$ and has finitely many edges, is the same as that for a ground state (as evident from  \eqref{Eq:exc1} and \eqref{Eq:exc2}) implying that the excitations are effectively localized in $R$ although the operator creating this excitation is not localized in $R$, but
    \item such excited states cannot be brought back to any ground state by the action of any operator which is supported within $R$.
\end{itemize}
Such considerations lead us to four superselection sectors which are as follows
\begin{itemize}
    \item the superselection sector $1$ as defined above,
    \item  the superselection sector of $e$ consisting of odd number of non-overlapping long string operators with one of their endpoints in $R$ (and their other endpoints at infinity), 
    \item  the superselection sector of $m$ consisting of odd number of non-overlapping long dual string operators with one of their endpoints in $R$ (and their other endpoints at infinity), and
    \item  the superselection sector of $d$ consisting of odd number of long string operators and odd number of long dual string operators with one of their endpoints in $R$ (and their other endpoints at infinity), and non-overlapping with each other.
\end{itemize}
We note the following:
\begin{itemize}
    \item Any state can be transformed to any other state within the same superselection sector by the action of an operator localized in $R$, e.g. the state created by the action of any odd number of long string operators ending at $R$ is equivalent to the state created by the action of a single long string operator ending at $R$ as each can be transformed to the other via the action of string operators localized within $R$ (the reduced density matrices of these two states on any finite bounded region of the lattice are identical).
    \item Two states belonging to two different superselection sectors cannot be transformed to each other via the action of any operator localized within $R$.
    \item All superselection sectors describe excitations effectively localized within $R$ because the reduced density matrix over any finite region $R'$, which does not overlap with $R$ and is at finite distance from $R$ is the same as that for a ground state although the operators creating the excitations are not necessarily localized in $R$.
\end{itemize}
Furthermore, we note that any state in the superselection sector $d$ is the product of an operator which creates a state in superselection sector $e$ with an operator which creates a state in superselection sector $m$. Here, the product commutes (although the operators themselves may not) since states are defined up to multiplication by phases. Given that the toric code has only four superselection sectors for any bounded region $R$, it is obvious that we can formalize a multiplication rule, a.k.a. fusion rule, defining operator equivalence classes and a fusion product under which the finite set of equivalence classes is closed. 

In what follows, we will give a formal definition of the operator equivalence classes and the fusion rules under which the set of equivalence classes is closed. Our definitions will reproduce the fusion rules of the toric code but the fusion product in the model of interest in this paper will be non-commutative and we will also need to include the equivalence class of the zero operator to achieve closure. Both of the latter features are novel and are necessitated by the presence of non-invertible operators.

\subsection{Generalization of the fusion category}
\paragraph{Spectral monoid:} In order to have a consistent definition of operator equivalence classes that correspond to superselection sectors, we need to first restrict ourselves to the set of operators which generate (excited) eigenstates of the Hamiltonian when they act on ground states, and which are closed under multiplication. Such a set of operators which generate excitations in a finite region $R$, forms a monoid, which we call the spectral monoid of $R$, and denote as 
$\mathcal{A}_R$. We define it more precisely as below.

Consider a set of operators $\mathcal{A}_R$ corresponding to a bounded sub-region $R$ of the lattice with the following properties:
\begin{itemize}
    \item \textbf{Spectrum generation}: For any $A\in \mathcal{A}_R$, $\ket{A} = A \mathcal{P}_0 \ket{S}$ is an energy eigenstate or vanishes for a standard state $\ket{S}$ where $\mathcal{P}_0$ is the projector to the ground state manifold.
    \item \textbf{Localization of excitations in $\mathbf{R}$}: For any finite subregion $R'$ which has no overlap with $R$ and is at finite distance from $R$, the reduced density matrix of any non-vanishing state ${\rm Tr}_{\overline{R'}}(\ket{A}\bra{A})$ corresponding to an $A\in \mathcal{A}_R$ as mentioned above, should coincide with that of a canonical ground state.\footnote{The choice of a specific ground state removes the logical operators from our discussion.}($\overline{R'}$ denotes the complement of $R'$.)
    \item \textbf{Closure}: For any $A, B\in \mathcal{A}_R$, $\ket{AB} = A B \mathcal{P}_0 \ket{S}$ is also an energy eigenstate or vanishes.
    \item \textbf{Completeness}: All energy eigenstates localized in $R$ can be generated by  a suitable $A\in\mathcal{A}_R$ acting on a ground state.
    \item \textbf{Minimality}: The set $\mathcal{A}_R$ should have a minimal number of generators. Furthermore, removal of any element from the generators of $\mathcal{A}_R$ should lead to a violation of one of the above properties.
\end{itemize}
$\mathcal{A}_R$ is a monoid since it includes the identity operator, and satisfies closure and associativity under matrix multiplication. We call the set $\mathcal{A}_R$ the spectral monoid of $R$. Following the discussion in the previous subsection, we note that although the excitations generated by operators in $\mathcal{A}_R$  are localized in $R$, the operators $A\in \mathcal{A}_R$ are not necessarily supported entirely within $R$. We also note that in a standard QFT with a unique ground state, such operators are the just the set of particle creation and annihilation operators when we consider the whole of space instead of a bounded and finite region $R$, with $\mathcal{P}_0$ essentially being the evolution in infinite Euclidean time so that it projects a generic state to the ground state. 

We want to emphasize that the spectral monoid as defined above is a highly non-trivial generalization of the monoid generated by the set of creation operators in quantum field theory. In the following subsection, we will show that the spectral monoid of our model satisfying the criteria mentioned above is \textit{unique}.

 \paragraph{Superselection sectors:} The superselection sectors can be derived from the equivalence classes of operators composing the spectral monoid $\mathcal{A}_R$ as follows. We say that $A \in \mathcal{A}_R$ and $B \in \mathcal{A}_R$ are equivalent, i.e. $A\equiv B$, if and only if there exists operators $C \in \mathcal{A}_R$ and $D \in \mathcal{A}_R$ \textit{both of which are entirely supported in $R$} and realize the following conditions
\begin{equation}
A \mathcal{P}_0 =C B \mathcal{P}_0, \quad {\rm and} \quad  B \mathcal{P}_0 = D A \mathcal{P}_0.
\label{Eq:defnequiv}
\end{equation}
We note that $C$ and $D$ are not necessarily inverses of each other, and in fact they can be non-invertible. It is easy to verify that the above indeed establishes an equivalence relation (satisfying reflexivity, symmetry and transitivity). We also identify the corresponding excitations $\ket{A}$ and $\ket{B}$ created by the \textit{equivalent} operators $A$ and $B$ to be in the same superselection sector, so that there exists a superselection sector corresponding to each equivalence class of operators in the spectral monoid.  

The above definition of equivalence class takes into account that there can exist energy eigenstates $\ket{A}$ and $\ket{B}$ with the excited vertices/faces in $R$ such that $\ket{B}$ can be obtained from $\ket{A}$ via the action of an operator supported in $R$ and acting on $\ket{B}$, but \textit{not} vice versa. In this case, $\ket{A}$ and $\ket{B}$ will not belong to the same superselection sector. One simple example of such a situation is an elementary pair of adjacent zero flux states in our model, which can be created by the action of $M_3$ supported on the shared edge acting on the ground state. This state cannot be brought back to the ground state via the action of any local operator belonging to the spectral monoid $\mathcal{A}_R$. Therefore, this eigenstate is not in the same superselection sector as the ground state although it can be created by the action of a local operator on the latter. 

It is also important to note that unlike in usual models the zero operator $(0)$ needs to be included for closure, since products of operators in $\mathcal{A}_R$ can vanish. In fact, one can readily check from Eq. \ref{Eq:M3} that $M_3^2 =0$ (we will discuss soon why the choice of $M_3$ as given by Eq. \ref{Eq:M3} follows from the requirement that it is an element of $\mathcal{A}_R$). {The equivalence class $0$ has a single operator which is $0$ and it corresponds to the zero vector in the Hilbert space of states.}\footnote{Note that there is no notion of addition within any superselection sector which is defined through the equivalence classes of operators in the spectral monoid. }
 
\paragraph{Fusion rules:} The fusion rules are defined as follows. Let $\mathcal{A}_{R}^{(i)}$ denote equivalence classes of the spectral monoid $\mathcal{A}_{R}$ with $i$ ranging from $1$ to $n$. With the symbols $\otimes$ and $\oplus$, we can write
\begin{equation}
    \mathcal{A}_{R}^{(i)} \otimes \mathcal{A}_{R}^{(j)} = f_{ij1} \mathcal{A}_{R}^{(1)} \oplus f_{ij2} \mathcal{A}_{R}^{(2)} \oplus \cdots 
 \oplus f_{ijn}\mathcal{A}_{R}^{(n)}, \quad \,{\rm for} \, k = 1,2,\cdots,n
\end{equation}
The symbol $\otimes$ on the left hand side implies we are considering arbitrary elements from both $\mathcal{A}_{R}^{(i)}$ and $\mathcal{A}_{R}^{(j)}$ and multiplying them, whereas $\oplus$ on the right hand side is an ``or'' symbol implying that the result of the multiplication is in one of the equivalence classes in $\mathcal{A}_R$. The fusion rules are specified by $f_{ijk}$, a.k.a. fusion coefficients, which are either $0$ or $1$ and denote the absence or presence of the various classes. Note that $f_{ijk}$ and $f_{jik}$ need not be equal, so the fusion product $\otimes$ could be non-commutative (we will explicitly see this in the context of our model in the following subsection). 

Note also that the fusion rules are typically associative, so we expect
\begin{equation}
     \sum_m f_{ijm}f_{mkl} = \sum_m f_{iml} f_{jkm}.
\end{equation}
Although matrix multiplication is associative and $\mathcal{A}_R$ is closed under matrix multiplication by construction, one can readily see that the associativity can fail if one does not generate sufficiently enough number of elements in the equivalence class corresponding to a non-vanishing fusion coefficient of any fusion product. Therefore, the above relation needs to checked explicitly. We have explicitly checked that associativity holds in our model.

A fusion rule is called non-Abelian if more than one fusion coefficients generated by a fusion product is non-zero. This terminology is derived from the fusion rules of non-Abelian anyons \cite{nayak2008non,Tong:2016kpv}.

It is possible that the fusion coefficients need to be defined more appropriately in order to capture the statistics of the excitations as in the usual constructions with anyons \cite{Fuchs:1993et,nayak2008non,Tong:2016kpv}. However, this involves cumbersome combinatorics with the result depending on the size and geometry on $R$ rather than the microscopic nature of the model itself. Therefore, we refrain from such a computation in this work. Nevertheless, we have explicitly verified that our fusion product is associative.

\subsection{Fusion rules of the model}\label{Sec:FRNI}

\paragraph{\textbf{Spectral monoid}:} We need to first show that we can define the spectral monoid $\mathcal{A}_R$ for any subregion $R$ on the lattice in our model, and then identify the equivalence classes (superselection sectors) and compute the fusion coefficients. 

To define a spectral monoid $\mathcal{A}_R$, we need to choose $M_3$ appropriately. The defining properties of $M_3$ used earlier implied $M_3\, M_1 = -M_3 = M_1\,M_3$ and that $M_3$ takes eigenvectors of $M_2$ with $\pm1$ eigenvalue to the eigenvector of $M_2$ with $0$ eigenvalue. Such an $M_3$ can also be of the form
\begin{equation}\label{Eq:M3n}
     M_3 = 
    \begin{pmatrix}
        0&1&-1\\
        1&0&0\\
        -1&0&0
    \end{pmatrix}
\end{equation}
instead of that given by Eq. \eqref{Eq:M3}. However, for the above choice $M_3^2 = I-M_1$, and therefore $M_3^2$ supported on a single edge does not produce an energy eigenstate when it acts on any ground state. This violates the closure property required of $\mathcal{A}_R$.  

We can demonstrate that $M_3$ defined by \eqref{Eq:M3} acting on a single edge is part of the spectral monoid $\mathcal{A}_R$. To see this let us define $M_4$ as below
\begin{equation}\label{Eq:M4}
     M_4 = M_3 \cdot M_2=
    \begin{pmatrix}
        0&1&1\\
        0&0&0\\
        0&0&0
    \end{pmatrix}
    \end{equation}
We find that 
\begin{align}\label{Eq:Prod1}
  & M_3\cdot M_3 =   M_3\cdot M_4 = M_4\cdot M_3 = M_4\cdot M_4  = 0, \nonumber\\
  & M_3 \cdot M_1 = M_1 \cdot M_3 = -M_3, \nonumber\\
  & M_4 \cdot M_1 = - M_1 \cdot M_4 = M_4, \nonumber\\
  & M_3 \cdot M_2 = M_4, \nonumber \\
  & M_4 \cdot M_2 = M_3, \nonumber \\
  & M_2 \cdot M_3 = M_2 \cdot M_4 = 0,
\end{align}
Note that $M_4$ anti-commutes with $M_1$ (while we recall that $M_3$ commutes with $M_1$). We can readily see that $M_4$ supported on a single edge will create zero flux face excitations on the two overlapping faces and also excite the two overlapping vertices to $-1$ eigenvalues of $A_v$. It is thus a hybrid excitation. Furthermore we recall that any ground state is invariant under the action of $M_2^2$ supported on a single edge, while $M_1^2 = I$. The product of $M_1$ and $M_2$ acting on a single edge creates a hybrid dyonic excitation $d$ in which the two adjacent faces have $-1$ eigenvalues of $B_f$ and the two adjacent vertices have $-1$ eigenvalues of $A_v$. The remaining products are readily deduced. We conclude that the set of operators generated by the action of $M_1$, $M_2$ and $M_3$ on a single edge indeed forms a monoid each element of which generates an energy eigenstate when it acts on a ground state. It is a complete set as we generate all possible combinations of the eigenvalues of $A_v$ for the two overlapping vertices and eigenvalues of $B_f$ for the two overlapping faces that are different from $1$, and it is also a minimal set as $M_1$, $M_2$ and $M_3$ form a minimal set of generators for the set of operators which can produce all such excitations by acting on a ground state. Finally, this set satisfies closure in the sense mentioned above.

Since the Hilbert space is a tensor product of those of the edges of the lattice, we can finally conclude that \textit{ the spectral monoid $\mathcal{A}_R$ is generated by $M_1$, $M_2$ and $M_3$ acting on the edges within $R$, any long string and any long dual string each with one end point within $R$ and another end point at infinity.} The inclusion of the long string and the long dual string into the set of generators is necessary for exactly the same reasons discussed previously in the context of the toric code (they produce a single electric/magnetic charge excitation entangled with the another electric/magnetic at infinity and this does not violate the requirement of effective localization of the excitation within $R$ as discussed before -- see Sec. \ref{Sec:Intutitive}). 

Note that the elements of the commutative stabilizer monoid localized in $R$ which preserve the parent ground state are part of the spectral monoid due to the requirement of closure as they are also generated by the actions of $M_1$ and $M_2$ on various edges of $R$. {However, the defining criterion of minimality implies that the annihilators $\{K_e\}$ which annihilates the ground states can be excluded from the spectral monoid, as they are not required for closure (which is evident from \eqref{Eq:Prod1}) and also we can generate the complete set of local excitations without including them. Nevertheless, it is obvious from \eqref{Eq:Prod1} that we need the zero ($0$) operator for closure.}

{We can also prove that the spectral monoid $\mathcal{A}_R$ of our model which satisfies the defining requirements is unique.} We have shown that the requirement of closure  makes the choice of $M_3$ unique if we consider it as a generator of the spectral monoid and acting on a single edge only. However, the choice of $M_{3,e}$ as a generator of the spectral monoid is still ambiguous because the pair of adjacent zero flux face excitation created by $M_{3,e}$ can also be created by another local operator.  As for instance, we can replace $M_{3,e}$ with $$\Phi_e = M_{3,e}P_{f_{e_1}}P_{f_{e_2}}$$where $f_{e_1}$ and $f_{e_2}$ are the two faces which overlap with the edge $e$. Since $\Phi_e \mathcal{P}_0 = M_{3,e}\mathcal{P}_0$, both $M_{3,e}$ and $\Phi_e$ create the same excited state (with zero fluxes at $f_{e_1}$ and $f_{e_2}$) acting on a ground state. However, we can check that $$\Phi_e \Phi_{e'}=0$$if the edges $e$ and $e'$ share any face (recall that $M_2M_3 =0$) although $$M_{3,e} M_{3,e'}\neq0.$$It follows that in order to create a fractonic cluster where both the edges $e$ and $e'$ have supported the action of $M_3$ (on a ground state), we need to necessarily include $M_{3,e}$ and $M_{3,e'}$ as generators (together they create the fractonic cluster state), {or another operator which creates this fractonic cluster state as a generator}. However, including both $M_{3,e}$ and $\Phi_e$, or including a separate generator for creation of a cluster state violates the criterion of minimality stated in the previous subsection. Therefore, we cannot replace $M_{3,e}$ with $\Phi_e$ as a generator of the spectral monoid. 

Similarly, if we can replace $M_{3,e}$ with another operator $\Phi'_{e}$, then it is necessary that $\Phi'_e \mathcal{P}_0 = M_{3,e}\mathcal{P}_0$ so that $\Phi'_{e}$ and $M_{3,e}$ create the same excited state acting on the ground state. Such a $\Phi'_{e}$ can differ from $M_{3,e}$ by attachments of operators in the stabilizer monoid, i.e. $\Phi'_{e} = M_{3,e}s$ where $s$ belongs to the stabilizer monoid ($s\mathcal{P}_0  = \mathcal{P}_0$). However, the stabilizer monoid is already a part of the spectral monoid (forming the equivalence class of the identity operator) and the generators of the stabilizer monoid are part of the generators of the spectral monoid. If $\Phi'_{e}$ and $M_{3,e}$ differ by attachment of invertible operators of the stabilizer monoid, then we obtain the same spectral monoid in both cases irrespective of whether $\Phi'_{e}$ or $M_{3,e}$ is considered as a generator. However, $\Phi_e$ discussed above differs from $M_{3,e}$ by attachment of non-invertible stabilizers, namely the elementary plaquette operators. In such cases, we can verify that we will always violate the requirement of minimality as in the case of $\Phi_e$. Repeating these observations for other generators, we can complete the proof that the spectral monoid $\mathcal{A}_R$ which satisfies our requirements is indeed unique. 

It is indeed non-trivial that we can generalize the set of creation operators of field theory in a unique way to the operators which create the locally excited states of our model from a ground state via the definition of the spectral monoid involving some simple requirements (closure, ability to generate complete spectrum, minimality, etc) that have been presented in the previous subsection.

{In fact the definition of the spectral monoid makes the notion of immobility of fractons precise. We note that the fractonic excitations created by the action of $M_3$ on an edge cannot be moved to another edge by the action of any operator belonging to the spectral monoid unlike the case of the long string operator ending on an edge which can be moved to another edge via the action of a short string. In our model, the fractonic excitations are thus completely immobile. A more general way to characterize the fractonic excitations (where the mobility can come with an energy cost) is via the fusion rules as discussed below.}

\paragraph{\textbf{Superselection sectors and fusion rules}:} If there are no zero flux face excitations in the region $R$, then the fusion rules are exactly that of the toric code. Aside from the identity, there are only three equivalence classes $e$ (electric charge), $m$ (magnetic charge) and $d$ (the dyon) discussed before which have the fusion rules,
\begin{align}\label{Eq:Fusion1}
    & e \otimes e = m \otimes m = d \otimes d = 1, \nonumber \\
    & e \otimes m = d = m \otimes e, \nonumber\\
    & m \otimes d = e = d \otimes m, \nonumber\\
    & d \otimes e = m  = e \otimes d.
\end{align}
The representative operators of $e$ and $m$ are a single long string (or an odd number of them) and a single long dual string (or an odd number of them), respectively, each with one endpoint in $R$ and another endpoint at infinity. Note that any even number of $e,m$ in $R$ is equivalent to the identity operator since such excited states can be created and transformed to the vacuum by acting appropriate $M_{1,2}$ strings, and this is captured by the fusion rules $ e \otimes e = m \otimes m =1$. The representative operator in $d$ is a product of one (or odd number of) long string(s) and one (or odd number of) long dual string(s) each with one endpoint in $R$ and another at infinity. {The identity equivalence class consists of the products of generators of the commutative stabilzer monoid which are localized within the region of interest $R$.} {Our definition of the spectral monoid not only leads to the same equivalence classes and fusion rules in the toric code sector of our model but also reproduces these in other similar stabilizer codes without elementary non-invertible stabilizers (it reduces to the construction of superselection sectors in \cite{Pai:2019fqg}\footnote{In \cite{Pai:2019fqg}, the authors did not need to construct a spectral monoid explicitly but the set of operators considered by them satisfy the defining properties of spectral monoid. In our case. we need an explicit construction of a spectral monoid because otherwise an equivalence relation and the notion of superselection sectors cannot be defined. Although fractonic excitations and their fusion products were considered in \cite{Pai:2019fqg}, they were not created by non-invertible operators acting on a parent ground state.}}).

When there are only two adjacent zero flux defect faces in $R$, we have the following additional equivalence classes:
\begin{enumerate}
    \item $q_{e}$ represented by the operator $M_3$ acting on a single edge $e$ in $R$,
    \item $q'_{e}$ represented by the product of $M_3$ acting on a single edge $e$ in $R$ and a long string operator ending in $R$ (the long string may or may not overlap on the edge $e$ -- see Fig. \ref{Fig:qprime}(a) and (b)).
\end{enumerate}
Note that both in Fig. \ref{Fig:qprime}(a) and Fig. \ref{Fig:qprime}(b), the electric charge excitation $e$ can be freely moved without changing equivalence classes via appropriate ($M_2$) strings localized in $R$ which do not intersect the edge where $M_3$ or $M_4$ acts. 

We note that $q_{e}$ and $q'_{e}$ are the only two additional equivalence classes of excitations localized on a single edge since the excitation created by the operator $M_4$ is equivalent to the $M_3$ excitation as illustrated in Fig. \ref{Fig:M4equiv}. Consider the operator $M_4 \mathcal{P}_0$. Let $v_{1,2}$ be the vertices labelling the edge where $M_4$ acts. As shown in Fig.~\ref{Fig:M4equiv}, we can multiply this from the left with an open $M_2$ string $E_{v_1,v_2}$ that ends at vertices $v_1$ and $v_2$ but doesn't act on the edge with $M_4$ to obtain
\begin{equation}
    E_{v_1,v_2} M_4 \mathcal{P}_0=  M_3 M_2 E_{v_1,v_2} \mathcal{P}_0 = M_3 \mathcal{P}_0,
\end{equation}
where we have used that fact that $E_{v_1,v_2}$ commutes with $M_4$ in the first equality since $E_{v_1,v_2}$ doesn't act on the same edge as $M_4$ and that $M_4 = M_3 M_2$. The second equality above uses the fact that $M_2 E_{v_1,v_2}$ is a contractible $M_2$ loop that can be absorbed into the ground state projector. It is also easy to see that starting from $M_3 \mathcal{P}_0$ we can run the same argument backwards to obtain 
\begin{equation}
    E_{v_1,v_2} M_3 \mathcal{P}_0 =  E_{v_1,v_2} M_3 M_2 E_{v_1,v_2} \mathcal{P}_0 = M_4 \mathcal{P}_0.
\end{equation} 
Therefore, it follows that the operators $M_3$ and $M_4$ acting on a single edge satisfy the equivalence relation defined in Eq. \eqref{Eq:defnequiv}. So the excitation created by the action of $M_4$ supported on a single edge on a parent ground state belongs to the $q_e$ equivalence class.

{It is obvious that we need to include the zero superselection sector as the zero operator should be in the spectral monoid for closure. The state(s) corresponding to this superselection sector is just the null state which is in any Hilbert space. The zero superselection sector contains \textit{only} the zero operator as the annihilators are not included in the spectral monoid as discussed before.}
\begin{figure}
    \centering
   \includegraphics[scale=0.5]{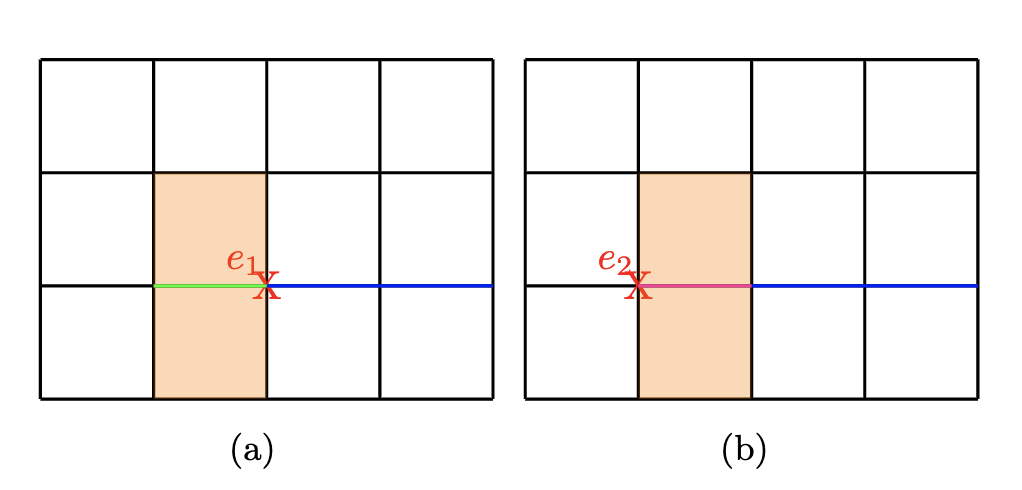}
    \caption{(a) An example of the $q'_e$ equivalence class where the long $M_2$ string (blue) doesn't intersect the edge $e$ where $M_3$ acts (green). (b) An example of the same $q'_e$ equivalence class where the long $M_2$ string intersects the edge $e$ where $M_4$ (magenta) acts. These two excitations are in the same equivalence class since one can transform from (a) to and from (b) using an open $M_2$ string connecting $e_{1,2}$ similar to the illustration in Fig.\ref{Fig:M4equiv}}
    \label{Fig:qprime}
\end{figure}
\begin{figure}
    \centering
    \includegraphics[scale=0.5]{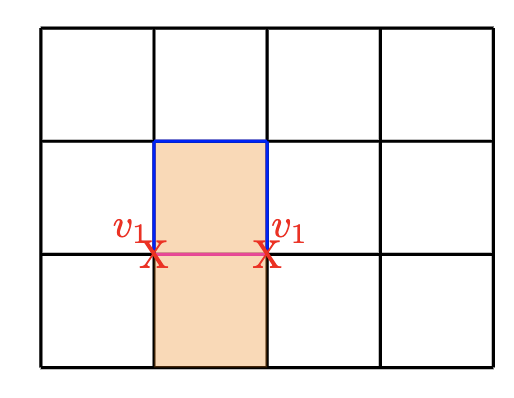}
    \caption{The action of $M_4 = M_3 M_2$ is denoted by the magenta edge. Consider $M_4 \mathcal{P}_0$ and as shown in the figure we multiply this from the left with an open $E_{v_1,v_2}$ string ending at $v_1$ and $v_2$. The $E_{v_1,v_2}$ string can be moved across $M_4$ and combines with the $M_2$ operator from $M_4$ to result in $M_3 M_{2c} \mathcal{P}_0$, where $M_{2c}$ is a contractible loop that can be absorbed into the ground state projector.}
    \label{Fig:M4equiv}
\end{figure}

Using Eq. \ref{Eq:Prod1}, we can obtain the following fusion rules involving $m$ and the defects $q_e$ and $q_e'$.
\begin{align}\label{Eq:Fusion2}
    & q_e \otimes q_e =  q'_e \otimes q'_e = 0, \nonumber \\
    & q_e \otimes m = m \otimes q_e = q_e, \nonumber\\
    & q'_e \otimes m = m \otimes q'_e = q'_e. \\
\end{align}
The last two fusion rules above which show that a defect can absorb a $m$ excitation imply the existence of magnetic monopoles outside any defect region as shown in Fig. \ref{Fig:monopole} and discussed before. Note that the mobility of $m$ excitations in the presence of defects\footnote{The monopole actually can move through the defect region and becomes invisible (absorbed) only when it coincides with an edge of the defect region where $M_3$ acts.} is reflected in the fact that the last two fusion rules above hold for any edge $e$. We also note that we need to include the zero operator (equivalence class) for closure.

The fusion rules between $e$ and the defects which can also be computed using Eq. \ref{Eq:Prod1} are more complex, and are as listed below.
\begin{align}\label{Eq:Fusion3}
    q_e\otimes e &= q'_e , \nonumber\\
    e\otimes q_e &= q'_e \oplus 0, \nonumber \\
    q'_e \otimes e&= q_e, \nonumber\\
    e \otimes q'_e &= q_e \oplus 0.
\end{align}
These fusion rules are both non-commutative and non-Abelian. 

The non-Abelian and non-commutative nature of the fusion rules can be explained as follows. Recall that the fusion product $\otimes$ involves multiplying all possible elements of the two equivalence classes being fused. On the right hand side we write down all the equivalence classes that appear in such a product with a $\oplus$ symbol indicating a logical 'or'. The non-commutative and non-Abelian nature of the fusion products $q_e\otimes e$ and $e\otimes q_e$ given above are illustrated in Fig. \ref{Fig:nonabelian}. As illustrated in this figure, we find that the non-Abelian properties of both these products arise from different choices of a representative long string operator $e$ giving different results depending on whether the long string does or does not intersect the edge which supports the action of $M_3$ in the class $q_e$. The non-commutativity of both the fusion products as also illustrated in this figure follows from $M_2\, M_3 = 0$ and $M_3\,M_2 = M_4$. Similar computations can be used to understand the other non-Abelian and non-commutative fusion rules listed above.

The appearance of the $0$ equivalence class on the right hand side illustrates that the $e$ excitations have restricted mobility along certain paths in the presence of the defects as shown in Fig. \ref{Fig:confcharge} and discussed before.

\begin{figure}[H]
    \centering
    \includegraphics[scale=0.56]{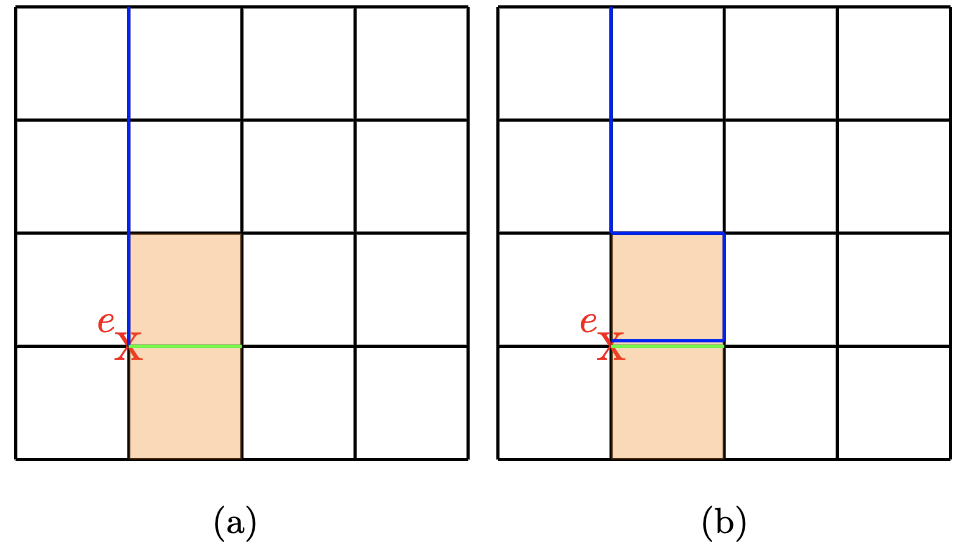}
   \caption{Illustration of the fusion product $e \otimes q_e$ and $q_e \otimes e$ depending on the order of multiplication. (a) Shows one representative element of the $e$ class multiplied with one of $q_e$ creating $q'_e$. Also shows an element of $q_e$ class multiplied with that of the $e$ class creating $q'_e$. (See Fig. \ref{Fig:qprime}(a) for a representative of $q'_e$ which is equivalent to the operator produced by these products.)  (b) Shows another representative element of the $e$ class that has an edge overlapping with the edge where $M_3$ acts. Since $M_2 M_3 =0$, the multiplication of this element with $q_e$ produces $0$. Since $M_3 M_2 =M_4$, which is equivalent to $M_3$, the multiplication of $q_e$ with this element produces $q'_e$. (See Fig. \ref{Fig:qprime}(b))}
     \label{Fig:nonabelian}
\end{figure}

The most striking non-commutative and non-Abelian elements of the fusion rules are actually those involving the identity sector (note that the toric code sector satisfies $e\otimes 1 =e = 1\otimes e$, $m\otimes 1 =m=1\otimes m$, $d\otimes 1 = d=1\otimes d$):\footnote{Since the left and right fusion products of $q_e$ with $1$ are different and can yield something different from $q_e$, etc., the symbol $1$ standing for the equivalence class of the identity operator can be confusing. Nevertheless, it is easy to see that both left and right fusion product of any equivalence class with $1$ yields back the equivalence class itself as one of the possibilities as in the standard multiplication with identity and $1\otimes 1 =1$.}
\begin{align}\label{Eq:Fusion4}
    1\otimes q_e &= q_e \oplus 0 , \nonumber\\
    q_e\otimes 1 &= q_e, \nonumber \\
    1\otimes q'_e &= q'_e \oplus 0 , \nonumber\\
    q'_e\otimes 1 &= q'_e.
\end{align}
This feature of non-triviality of multiplication with $1$ is a consequence of the contractible $M_2$ loop being equivalent to the identity operator as it is an element of the stabilizer monoid. The first two fusion rules above are illustrated in Fig. \ref{Fig:identityfusion}. Depending on whether the contractible $M_2$ loop intersects the edge $e$ or not, we get different results on the right hand side. Thus, the presence of non-invertible symmetries gives rise to this striking feature that products with the identity sector is both non-Abelian and non-commutative. As shown in Sec. \ref{Sec:Detect}, we can distinguish two fractonic excitations with identical zero flux faces only because contractible $M_2$ loops are part of the stabilizer monoid. It is interesting to see that the same $M_2$ loops are also responsible for the feature of non-triviality of the fusion product with identity. 

Note that generalized symmetries usually form a non-trivial fusion category instead of a simple mathematical structure like a group \cite{Schafer-Nameki:2023jdn,Shao:2023gho}. In our model, although the generalized symmetries themselves form a monoid, the non-trivial stucture of the stabilizer monoid (especially the feature that we need to add more generators than the elementary stabilizers) gives rise to the non-triviality of the fusion product with identity. 

\begin{figure}[H]
    \centering
    \includegraphics[scale=0.5]{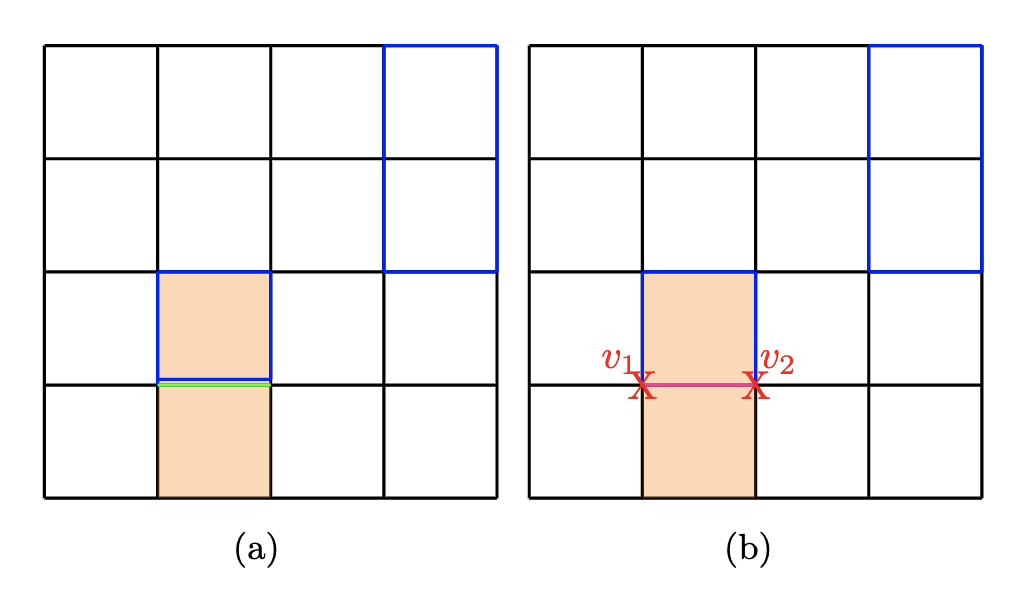}
    \caption{Illustration of the fusion product $1 \otimes q_e$ and $q_e \otimes 1$ depending on the order of multiplication. (a) Illustration of $1\otimes q_e$. Shows one representative element of the $1$ class, a contractible $M_2$ loop (blue) intersecting with $M_3$ (green) that creates $q_e$. Since $M_2 M_3 =0$ this element of the equivalence class gives $0$ after multiplication. Another element of the $1$ class is the other contractible $M_2$ loop in the figure that doesn't intersect $M_3$. This can be simply moved across $M_3$ and absorbed into the ground state projector to obtain $q_e$. (b) Illustration of $q_e \otimes 1$. If the order of multiplication in (a) is reversed we obtain a product of $M_4 = M_3 M_2$ (magenta) and a open $M_2$ string $E_{v_1,v_2}$ consisting of the remaining part of the loop, which is equivalent to $q_e$ (see Fig.\ref{Fig:M4equiv}). The other element of the $1$ class is the other contractible loop in the figure. This can simply be absorbed into the ground state projector to obtain $q_e$.}
    \label{Fig:identityfusion}
\end{figure}
{Furthermore, one can also check the fusion rules involving the identity sector are consistent with associativity. Actually we can also determine $1\otimes q_e$, $q_e\otimes 1$, etc. via associativity, and using $e\otimes e =1$ and \eqref{Eq:Fusion3}, since $$1\otimes q_e = e\otimes (e\otimes q_e) = e\otimes (q'_e\oplus 0)= q_e \oplus 0,$$$$q_e\otimes 1 = (q_e\otimes e)\otimes e = (q'_e)\otimes e =q_e,$$etc. We can also check that the remaining fusion rules involving defects where only a single edge supports the action of $M_3$ or $M_4$, along with those listed above, are associative, and all of these remaining ones can be determined via associativity as well, e.g. using $e\otimes m = m\otimes e =d$ and \eqref{Eq:Fusion3} we obtain $$d\otimes q_e = e\otimes(m\otimes q_e) = m\otimes(e\otimes q_e) = q'_e\oplus 0,$$etc.}

Our fusion rules can mathematically be best understood as a \textit{semi-category} (a.k.a \textit{non-unital category}) \cite{Mitchell11972,Moens2002}\footnote{See also: https://ncatlab.org/nlab/show/semicategory.} where the composition of morphisms between objects is associative but there need not exist any identity morphism (this category generalizes semi-groups and semi-rings). It would be interesting to explore how such structures can appear also in quantum field theories.

The fusion of two defect excitations $q_e$ and $q_{e'}$ corresponding to different edges $e$ and $e'$ creates a new equivalence class:
\begin{equation}
    q_e \otimes q_{e'} = q_{e'} \otimes q_e =q_{ee'}
\end{equation}
This proliferation of equivalence classes, i.e. scaling of the number of superselection sectors with the size of the system, is characteristic of fractonic models and captures the immobile nature of the corresponding fractonic excitations \cite{Pai:2019fqg}. However, we can readily see that some of the internal states of a larger defect configuration generated via actions of $M_1$ in the internal edges as discussed before, will not form a new equivalence class. For a larger defect configuration, the equivalence classes will denote inequivalent ways of producing the zero flux configurations solely via action of $M_3$ on the internal edges. 

{In typical systems, the mobile (anyonic) excitations constitute a finite number of equivalence classes whose fusion products are commutative and close amongst themselves \cite{Tong:2016kpv}. If the model has fractonic equivalence classes, then the number of equivalence classes grow with the size of the region. Therefore, removing the set of fractonic equivalence classes leads to a commutative truncation of the fusion rule such that the number of elements do not scale with the size of the region of interest. This can be understood as a more general way of defining equivalence classes involving fractonic excitations. In fact, as discussed in \cite{Pai:2019fqg}, this definition of fractonic equivalence classes can include not only fully immobile fractons, but also fractons whose mobility via local operations can incur an energy cost. It would be interesting to investigate whether fusion rules (as defined here) can also distinguish between different types of fractonic excitations.}

Finally, we note that in presence of larger defect configurations, one can have new equivalence classes, such as one involving even number of electric charges, due to the restricted mobility of electric charge excitations in the presence of the defects. For an illustration see Fig. \ref{Fig:ECed} where a string of $M_3$ operators creates a wall of zero flux faces that partition the region $R$ (shaded in gray) such that no string can pass from one half to the other without intersecting an edge where $M_3$ acts. Therefore, adding a pair of long string operators creating an electric charge on each half afterwards creates a new equivalence class since these electric charges cannot be neutralized by a string operator located within $R$ without acting on one of the edges where $M_3$ acts (the latter would annihilate the state as $M_2 M_3 =0$).

\paragraph{\textbf{Summary}:} To summarize, we have shown that all locally excited states of our model can be created from a parent ground state by a unique set of operators which form the spectral monoid satisfying some simple requirements like closure, ability to generate the complete spectrum, minimality of generators, etc., and generalizing the set of creation operators of a perturbative quantum field theory. Furthermore, the operators of the spectral monoid can be classified into equivalence classes with the associated states forming superselection sectors.  The fusion products of our model can be readily drived from these equivalence classes. These fusion product is associative, but both non-commutative and non-Abelian. Strikingly, the product with identity is itself non-commutative. The fusion rules encapsulate the properties of the excited states especially (im)mobility or partial mobility, and also other features like the ability of fractonic defects to absorb the magnetic charges. A subset of the equivalence classes representing fully mobile excitations is identical to that of the toric code with the usual fusion rules. As mentioned above, our fusion rules are an example of a semi-category (a.k.a. non-unital category) in which the composition of morphisms between objects is associative but a proper identity morphism need not exist.

\begin{figure}[H]
    \centering
    \includegraphics[scale=0.56]{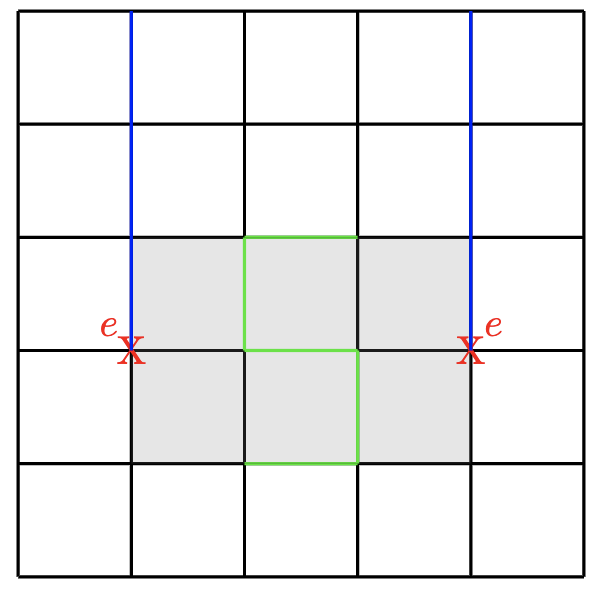}
    \caption{Another nontrivial equivalence class. The gray shaded region including the edges on its boundary denotes the region of interest $R$. Note that the pair of electric charges cannot be converted to the ground state via the action of $M_2$ strings supported only within the region $R$. Any such $M_2$ string will have to intersect at least one green $M_3$ edge leading to the annihilation of the state.}
 \label{Fig:ECed}
\end{figure}

\paragraph{\textbf{Discussions}:} The operators of the spectral monoid are the generalizations of the creation operators of the Fock space of a free quantum field theory. Therefore, the spectral monoid fully characterizes the mutual statistics of the excitations. The electric and magnetic charge excitations commute amongst themselves but mutually anti-commute if they overlap with each other (since $M_1$ and $M_2$ anti-commute) as in the toric code. The fractonic excitation created by $M_{3}$ supported on a edge behaves partly as a fermion since $M_3^2 =0$ (satisfying Pauli exclusion). However, the products of two $M_3$ acting on two different edges commute, so the fractonic excitations localized on different edges behave as bosonic excitations as the corresponding wavefunction is symmetric in real space. These features are partly captured in the fusion products such as $q_e$ squares to zero and $q_e \otimes q_{e'} = q_{e'} \otimes q_e$. 

The most important statistical feature of the excitations comes from the products $M_2 M_3 = 0$ and $M_3 M_2 = M_4$. These imply that an ordering of operators in which $M_2$ acts after $M_3$ to create an excited state is simply not possible. The non-commutative fusion product of $q_e$ ad $e$ does bring out this feature of the hybrid excitation. 

A concrete question which can be asked is whether the fusion product contains complete information about the degeneracy of the excited states belonging to a specific superselection sector arising from the fusion product. For anyonic systems, this is indeed the case as can be exemplified by the feature that the quantum dimension of an anyonic excitation can be extracted from the repeated fusion products of the anyon with itself \cite{Tong:2016kpv}. To generalize this feature in our model, the fusion coefficients may need to take appropriate values other than the simple values $0$ and $1$ we have considered here. As mentioned before, this is a difficult combinatoric problem which we leave for the future. It could be possible that we need to take a new approach to capture statistics and braiding in models with fractonic excitations -- see \cite{Shirley:2018nhn} for a recent discussion. {We need to investigate also the related consistency conditions such as pentagon and hexagon identities which arise from the relation between braiding and the fusion rules.} In any case, it is worth emphasizing that our fusion rules defined with fusion coefficients $0$ and $1$ satisfy associativity.

One important element to note is that the superselection sectors characterize the different types of excitations in our model and are not commensurate with the number of ground states unlike the toric code. Intuitively the Wilson lines (logical operators) which create the other ground states from a parent one are also world-lines where a quasi-particle and its anti-partner can be pair created from the parent ground state and annihilated to produce another ground state. However, when there are immobile excitations, the number of different types of excitations need not scale with the number of ground states. Nevertheless, in typical fracton models, the number of ground states, scale with the system size and so does the number of superselection sectors \cite{Nandkishore:2018sel,Pretko:2020cko}. Although in our model, this is not the case, as discussed in Sec. \ref{Sec:Decon}, a deformation of our model has a fracton liquid phase with deconfined fractons and also with the number of ground states scaling with the system size. It would be interesting to understand the fusion rules of the fracton liquid phase. As should be clear from our construction, the equivalence classes and superselection sectors depends explicitly on the projector to the manifold of ground states. Therefore, different phases which can result from the change of a coupling constant, should have different fusion rules. Similarly, it would be interesting to study the fusion rules of the phase discussed in Sec. \ref{Sec:Decon} where the dual string operators have condensed.

It is also an interesting question to ask whether the fusion rules can be used to distinguish different phases containing confined or deconfined fractons which can result from Hamiltonians of our type containing non-invertible stabilizers. The quotienting of the equivalence classes by translation as done recently in \cite{Pai:2019fqg} will be a complicated exercise but could be needed to distinguish between fractonic models of this type. We leave this investigation for the future.

Finally, we remark that a simple topological quantum field theory (TQFT) cannot capture the superselection sectors of our model unlike the case of the $\mathbb{Z}_2$ gauge theory \cite{ReadSachdev,WenSpinLiquid,Bais:1991pe,Maldacena:2001ss} where the Wilson loops capture the superselection sectors and the fusion rules of the toric code. This is simply because the number of superselection sectors depend on the lattice in a specific way. Consider a fractonic defect region on the lattice. This can be created from a parent ground state using different choices of edges supporting the action of $M_3$. Each of these choices corresponds to a distinct superselection sector. So the number of superselection sectors corresponding to a fractonic defect region depends on the details of the lattice structure in that region. It would be interesting to see if a generalized TQFT (perhaps with elements of emergent gravity which can process relevant geometric information) can be constructed to reproduce the superselection sectors and the fusion rules ({see \cite{Kim1,Aasen:2020zru} for related discussions.})

}

\section{Outroduction}\label{Sec:Outro}

In this work, we have introduced a model with non-invertible stabilizers in which novel confined fractonic excitations which can modify the properties of the deconfined mobile exciations exist. The ground states degeneracy of the model is topological like that of the toric code and also a subset of the exciations are isomorphic to that of the toric code. Furthermore, we have shown that such models can be analyzed via a \textit{unique} spectral monoid which generalizes the notion of creation operators in a quantum field theory that create excited states from a parent ground state, but with the excitations in our case localized to an arbitrarily chosen finite region. We have shown how superselection sectors and fusion rules can be derived from the unique spectral monoid. These fusion rules are associative, but both non-commutative and non-Abelian, and form a semi-category (a.k.a. non-unital category) where a proper identity morphism does not exist. The fusion rules capture many physical properties of the excitations.

The model studied in this paper is part of a general class of stabilizer codes with a qutrit at each edge and with non-invertible symmetries. One such class is obtained from the family 
\begin{equation}\label{Eq:M1M2n}
    M_1 \equiv 
    \begin{pmatrix}
        r&0&0\\
        0&0&1\\
        0&1&0
    \end{pmatrix} = r \bigoplus X, \quad
    M_2 \equiv 
    \begin{pmatrix}
        0&0&0\\
        0&1&0\\
        0&0&-1
    \end{pmatrix} = 0 \bigoplus Z,
\end{equation}
with $r$ being a real number. (The case $r=-1$ corresponds to the present model.) We readily see that $M_1$ and $M_2$ anti-commute for arbitrary $r$. We can define the model in terms of elementary stabilizers, namely vertex operators $A_v$, each of which corresponds to a vertex $v$ and is the product of $M_1$ acting on each edge converging at $v$, and plaquette operators $B_f$, each of which corresponds to a face $f$ and is the product of $M_2$ acting on each edge bounding  $f$. These elementary stabilizers commute with each other. The Hamiltonian is the sum of all the stabilizers up to an overall negative multiplicative constant. 

It is easy to see that the toric code ground states, which are in the subspace obtained from the tensor product of the lower block two-dimensional subspaces of each edge, will be the ground states of these models unless $r$ is a sufficiently large number. If $r$ is sufficiently large, then the energy is minimized by the fully unentangled state
\begin{equation}
    \ket{G} = \bigotimes \prod_e \begin{pmatrix}
        1\\0\\0
    \end{pmatrix},
\end{equation}
since in this case the vertex operator is more relevant at low energy and the lowest energy state should maximise the eigenvlues of each $A_v$. However, for intermediate values of $r$, energy minimization is a complex problem which would depend on the lattice. 

Furthermore, one can check that $M_3$ (given by \eqref{Eq:M3}) supported on any edge $e$ and acting on the ground states generates fractonic confined defects. We postpone the analysis of such models to a future publication. It would be of course first important to see that whether our construction leads to a  unique spectral monoid for such models, and then study the nature of the derived fusion rules.



One can usually generalize stabilizer code models to obtain phases of a discrete gauge theory. A generalization of the toric code gives phases where the electric or magnetic charges are confined while going through the quantum critical topological point described by the toric code itself \cite{Wegner,RevModPhysKogut} (see also \cite{ReadSachdev,WenSpinLiquid,Bais:1991pe,Maldacena:2001ss}). These phases can be captured by a $\mathbb{Z}_2$ gauge theory with confined/Higgsed phases included \cite{FradkinShenker}. More recently, it has been shown that higher group gauge theories can describe phases of more general stabilizer codes \cite{baez2005higher, ibieta2020topological, de2017topological, bullivant2017topological, bullivant2020higher,Barkeshli:2022edm,Pace:2022cnh}. A natural question therefore is whether one can construct a quantum field theory which reproduces the phase described by the model studied in this paper and other phases such as deconfined fracton liquids, etc. obtained from variations of our model as discussed in Sec. \ref{Sec:Decon}.

We readily realize that one can create an even larger family of such models featuring non-invertible symmetries and non-invertible defects with a qudit, i.e. a $d$ level system at each edge of the lattice by choosing $M_1$ and $M_2$ (which would be $d\times d$ matrices) appropriately.\footnote{See \cite{Padmanabhan:2022ajc} for models with two qubits on each edge and a groupoid extension of the toric code. This model also features fractons.} 

It is not obvious that the construction of the unique spectral monoid and fusion category which is a semi-category (a.k.a. non-unital category) in this work readily generalizes to all such lattice models with $d\geq3$. It is also not obvious that the fusion product is necessarily associative (generalizations of semi-category without associativity in the composition of morphisms exists \cite{Salvatore2013}). It would be of course important to understand how such semi-categories or their generalizations can arise in quantum field theories which can describe the continuum limit of these lattice models.

Furthermore, we need to also establish a connection between fusion rules and statistics with appropriate ways to define the fusion coefficients, and also a notion of braiding including fractonic excitations (see \cite{Shirley:2018nhn} for related work). Finally, we would need to how renormalization group transformations can classify these models in terms of symmetries and fusion rules (see \cite{Pai:2019fqg} for such a discussion in the context of fracton models). 

 Another set of fundamental questions is as follows. It is known that anyon fusion rules in two-dimensional lattice theories can be mapped to the fusion rules obtained from the operator product expansion (OPE) of local operators in rational CFTs \cite{Nayak_2008} (see \cite{Buican:2017rxc} for a recent development). Therefore, it is natural to ask the following questions.
\begin{itemize}
    \item Can the fusion rules of our lattice model and those of the generalizations described above be reproduced from the fusion rules derived from the OPEs of a scale-invariant continuum theory? 
    \item The requirement that the conformal blocks of a CFT provide a representation of the mapping class (braid) group of a Riemann surface gives non-trivial constraints on the fusion rules and also imply profound relations such as the Verlinde formula relating fusion coefficients with the modular transformations \cite{Verlinde:1988sn, Moore:1988uz}. Can similar fundamental constraints and relations be derived for our generalized fusion rules? 
    \item Recently profound relations have been understood between entanglement measures and the fusion rules both in CFTs \cite{He:2014mwa} and in lattice models \cite{Shi:2018bfb,Shi:2019mlt,Shi:2019ngw}. Can we connect entanglement with fusion rules both in our lattice model and its generalizations, and also in the related continuum field theories via a general axiomatic paradigm? 
\end{itemize}

It would be also interesting to see if one can construct a family of models where $d$ (the level of the qudit living on an edge of the lattice) can be varied and where we can take the limit $d\rightarrow\infty$. Analyzing this limit with an infinite-dimensional Hilbert space on every edge is similar to taking the large $N$ limit in $\mathbb{Z}_N$ gauge theory. We can expect that this limit can be described by a large $N$ type quantum field theory.

 We conclude with the observation that stabilizer codes have been useful in constructing models of holographic bulk reconstruction (see \cite{Kibe:2021gtw} for a recent review), and in this context fracton models (see \cite{Yan:2018nco} for an example) with non-invertible symmetries may give new insights into how gravity and spacetime can be reconstructed from degrees of freedom at the boundary.

\acknowledgments{We thank Abhishek Chowdhury, Benoit Doucot, Debashis Ghoshal, Suresh Govindarajan, Prabha Mandayam, Giuseppe Policastro, P Ramadevi and Pratik Roy for many stimulating discussions, and Arindam Lala, Archana Maji and Pratik Roy for initial collaboration in the early stages of this project. TK ackowledges support from the PMRF scheme of the Government of India. TK has also been partly supported by the Center of Excellence initiative of the Ministry of Education of India. AM acknowledges support from Fondecyt grant 1240955.}

\appendix
\section{Trace computations}\label{App:Trace}
\subsection{Isolated defect faces are disallowed}
\begin{figure}
    \centering
    \includegraphics[scale=0.56]{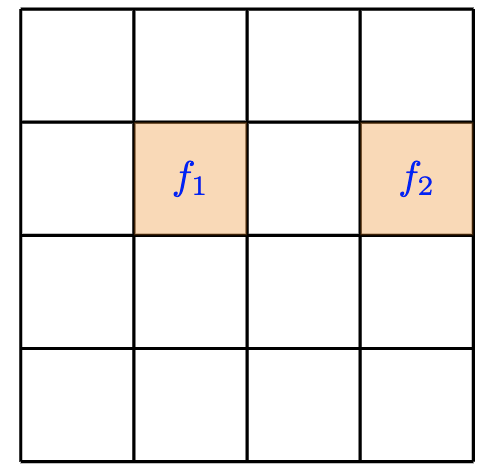}
    \caption{The impossible two face excitation configuration on separated faces $f_1$ and $f_2$. }
 \label{Fig:append1}
\end{figure}

Let us consider the following projector ($\mathcal{P}_2 = \mathcal{P}^d_{f_1,f_2}$)
\begin{equation}
    \mathcal{P}_2 = \left(I - B_{{f}_{1}}^2\right) \left(I - B_{{f}_{2}}^2\right)\prod_{v} \frac{1}{2} \left(I + A_{v}\right) \prod_{f\neq {f}_{1,2}} \frac{1}{2}\left(B_{f} + B_{f}^2\right).
\end{equation}
corresponding to states with the $f_{1,2}$ faces excited to the $0$ eigenstate. We will assume that the faces $f_{1,2}$ do not share any edge as shown in Fig.\ref{Fig:append1}.
This projector can be expanded as 
\begin{equation}
     \mathcal{P}_2 = \frac{1}{2^{|v|}2^{|f|-2}} \left(I + \prod_v A_v\right) \left(\prod_f B_f^2 - \prod_{f\neq {f}_1}B_f^2 - \prod_{f\neq {f}_2}B_f^2 + \prod_{f\neq {f}_{1,2}}B_f^2\right) + \text{traceless terms}.
\end{equation}
Note that terms such as $\left(I-B_{f_1}^2\right)\prod_{f \neq f_1} B_f$ are traceless. Since $\prod_v A_v = I$ on a closed surface and
\begin{equation}
    \prod_f B_f^2  = \prod_{f\neq \tilde{f}_1}B_f^2 =  \prod_{f\neq \tilde{f}_2}B_f^2 =\prod_{f\neq \tilde{f}_{1,2}}B_f^2 = \prod_e M_2^2,
\end{equation}
it is easy to see that
\begin{equation}
    {\rm Tr}(\mathcal{P}_2) = 0.
\end{equation}
Thus states with two separated zero eigenvalue face excitations are disallowed. Similar computations can be used to show that any configuration with a disconnected face defect is disallowed.

\subsection{Degeneracy of two zero flux face defects}
\begin{figure}
    \centering
    \includegraphics[scale=0.56]{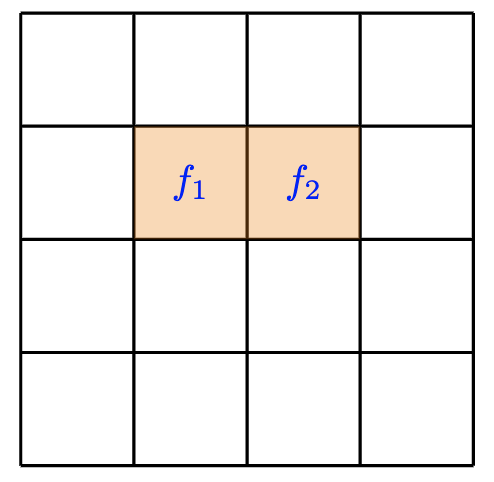}
   \caption{A reproduction of a part of Fig. \ref{Fig:exc3} from the main text showing two face excitations on connected faces ${f}_{1,2}$.}
 \label{Fig:append2}
\end{figure}

The case when we have two connected face excitations $f_{1,2}$ is shown in Fig. \ref{Fig:append2}. 
The projector in this case is ($\tilde{\mathcal{P}}_2 = \mathcal{P}^d_{f_1,f_2}$)
\begin{equation}
    \tilde{\mathcal{P}}_2 =  \left(I - B_{{f}_{1}}^2\right)  \left(I - B_{{f}_{2}}^2\right)\prod_{v} \frac{1}{2} \left(I + A_{v}\right) \prod_{f\neq {f}_{1,2}} \frac{1}{2}\left(B_{f} + B_{f}^2\right),
\end{equation}
which can be expanded as
\begin{equation}
     \tilde{\mathcal{P}}_2 = \frac{1}{2^{|v|}2^{|f|-2}} \left(I + \prod_v A_v\right) \left(\prod_f B_f^2 - \prod_{f\neq {f}_1}B_f^2 - \prod_{f\neq {f}_2}B_f^2 + \prod_{f\neq {f}_{1,2}}B_f^2\right) + \text{traceless terms}.
\end{equation}
Note that 
\begin{equation}
    \prod_f B_f^2 = \prod_{f\neq {f}_1}B_f^2 = \prod_{f\neq {f}_2}B_f^2 = \prod_e M_2^2,
\end{equation}
however the last term $\prod_{f\neq {f}_{1,2}}B_f^2$ corresponds to $M_2^2$ acting on all edges except the edge shared by $f_{1,2}$, say $e_{1,2}$, where we have a $3\times3$ identity operator. Thus,
\begin{equation}
    \prod_{f\neq {f}_{1,2}}B_f^2 = I_{e_{1,2}}\prod_{e\neq e_{1,2}} M_2^2.
\end{equation}
We therefore obtain that
\begin{equation}
    {\rm Tr}(P_2) = \frac{2}{2^{|v|}2^{|f|-2}} \left(2^{|e|} -2^{|e|} -2^{|e|} +3\times 2^{|e|-1}  \right) = \frac{2^{|e|}}{2^{|v|}2^{|f|-2}} = 2^{2g}.
\end{equation}

\subsection{Degeneracy of three zero flux face defects}
\begin{figure}
    \centering
    \includegraphics[scale=0.56]{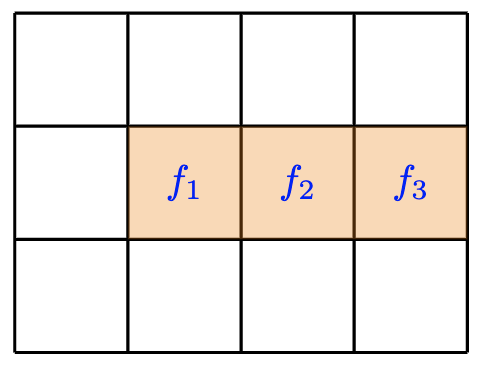}
    \caption{A reproduction of Fig. \ref{Fig:exc3a} from the main text showing three faces $f_{1,2,3}$ are excited to the zero eigenstate.}
 \label{Fig:append3}
\end{figure}

The trace of the projector for a configuration with three connected defect faces as shown in Fig. \ref{Fig:append3} can be computed as follows.
Let us label the edge connecting $f_{1,2}$ as $e_1$ and the edge connecting $f_{2,3}$ as $e_2$. The projector onto this subspace is then ($\mathcal{P}_3 = \mathcal{P}^d_{f_1,f_2,f_3}$)
\begin{equation}
    \mathcal{P}_3= \prod_{i=1,2,3}\left(I - B_{{f}_{i}}^2\right)\prod_{v} \frac{1}{2} \left(I + A_{v}\right) \prod_{f\neq {f}_{1,2,3}} \frac{1}{2}\left(B_{f} + B_{f}^2\right),
\end{equation}
which can be expanded as
\begin{multline}
 \mathcal{P}_3 = \left(-\prod_f B_f^2 + \prod_{f\neq {f}_1}B_f^2 + \prod_{f\neq {f}_2}B_f^2+ \prod_{f\neq {f}_3}B_f^2 - \prod_{f\neq {f}_{1,2}}B_f^2 -\prod_{f\neq {f}_{2,3}}B_f^2-\prod_{f\neq {f}_{1,3}}B_f^2 + \prod_{f\neq f_{1,2,3}}B_f^2 \right) \\ \times \frac{1}{2^{|v|}2^{|f|-3}} \left(I + \prod_v A_v\right) + \text{traceless terms}.
\end{multline}

\begin{multline}
 \mathcal{P}_3 = \left(-\prod_e M_2^2 + \prod_e M_2^2 + \prod_e M_2^2+ \prod_e M_2^2 - I_{e_1}\prod_{e\neq e_1} M_2^2 -I_{e_2}\prod_{e\neq e_2} M_2^2-\prod_e M_2^2 + I_{e_1} I_{e_2}\prod_{e\neq e_{1,2}} M_2^2 \right) \\ \times \frac{1}{2^{|v|}2^{|f|-3}} \left(I + \prod_v A_v\right) + \text{traceless terms},
\end{multline}
where $I$ denotes the $3\times3$ identity operator. Computing the trace of this projector we get
\begin{equation}
    {\rm Tr}(\mathcal{P}_3) = \frac{2}{2^{|v|}2^{|f|-3}} \left(-2^{|e|} +3\times 2^{|e|} -3 \times 2 \times 2^{|e|-1} - 2^{|e|} + 3^2\times 2^{|e|-2}\right),
\end{equation}
where we have carefully accounted for which of the internal edges $e_{1,2}$ have an identity and which have a $M_2^2$. Simplifying the expression for the trace, we get
\begin{equation}
    {\rm Tr}(P_3) = 2^{2g}.
\end{equation}

\subsection{Degeneracy of four zero flux face defects}
\begin{figure}
    \centering
    \includegraphics[scale=0.56]{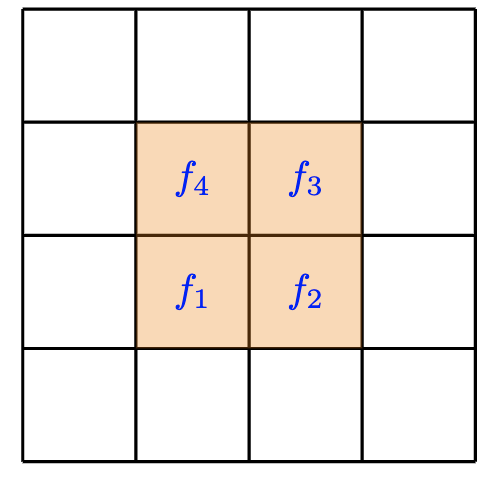}
    \caption{A reproduction of Fig. \ref{Fig:exc4b} from the main text showing four faces $f_{1,2,3,4}$ are excited to the zero eigenstate. Note that we do not indicate the various internal states in this figure.}
 \label{Fig:append4}
\end{figure}

The trace of the projector onto the subspace with four defect faces arranged in a square, as shown in Fig.\ref{Fig:append4}, can be computed as follows.
The projector onto this subspace is denoted by $\mathcal{P}_4$ and is given by ($\mathcal{P}_4 = \mathcal{P}^d_{f_1,f_2,f_3,f_4}$)
\begin{equation}
    \mathcal{P}_4= \prod_{i=1,2,3,4}\left(I - B_{{f}_{i}}^2\right)\prod_{v} \frac{1}{2} \left(I + A_{v}\right) \prod_{f\neq {f}_{1,2,3,4}} \frac{1}{2}\left(B_{f} + B_{f}^2\right),
\end{equation}
expanding out the projector we get
\begin{multline}
    \mathcal{P}_4 = \Biggl(\prod_f B_f^2 - \prod_{f\neq {f}_1}B_f^2 - \prod_{f\neq {f}_2}B_f^2- \prod_{f\neq {f}_3}B_f^2 -\prod_{f\neq {f}_4}B_f^2 \\+ \prod_{f\neq {f}_{1,2}}B_f^2 +\prod_{f\neq {f}_{1,3}}B_f^2+\prod_{f\neq {f}_{1,4}}B_f^2 +\prod_{f\neq {f}_{2,3}}B_f^2 + \prod_{f\neq {f}_{2,4}}B_f^2+\prod_{f\neq {f}_{3,4}}B_f^2\\ -\prod_{f\neq f_{1,2,3}}B_f^2 -\prod_{f\neq f_{1,2,4}}B_f^2-\prod_{f\neq f_{1,3,4}}B_f^2-\prod_{f\neq f_{2,3,4}}B_f^2 + \prod_{f\neq f_{1,2,3,4}}B_f^2 \Biggr) \\ \times \frac{1}{2^{|v|} 2^{|f|-4}} \left(I+\prod_v A_v + A_{v_0} + \prod_{v\neq v_0} A_v\right) +\text{ traceless terms }.
\end{multline}
Note that in this case we need to be careful to include the non-trivial contributions from the vertex operator on the vertex $v_0$ that is inside the defect region. Since the model is placed on a torus we have the constraints $\prod_v A_v = I$ and $\prod_{v\neq v_0} A_v = A_{v_0}$. The vertex operator $A_{v_0}$ will contribute to the trace only if all the internal edges on which it acts have no $M_2$ or $M_2^2$ operator since ${\rm Tr}(M_1 M_2) = {\rm Tr}(M_1 M_2^2) = 0$. When evaluating the trace of the above expression one must also recall that ${\rm Tr}(A_{v_0}) = 1$. We can further simplify the projector as
\begin{multline}
    \mathcal{P}_4 = \Biggl(\prod_f B_f^2 - \prod_{f\neq {f}_1}B_f^2 - \prod_{f\neq {f}_2}B_f^2- \prod_{f\neq {f}_3}B_f^2 -\prod_{f\neq {f}_4}B_f^2 \\+ \prod_{f\neq {f}_{1,2}}B_f^2 +\prod_{f\neq {f}_{1,3}}B_f^2+\prod_{f\neq {f}_{1,4}}B_f^2 +\prod_{f\neq {f}_{2,3}}B_f^2 + \prod_{f\neq {f}_{2,4}}B_f^2+\prod_{f\neq {f}_{3,4}}B_f^2\\ -\prod_{f\neq f_{1,2,3}}B_f^2 -\prod_{f\neq f_{1,2,4}}B_f^2-\prod_{f\neq f_{1,3,4}}B_f^2-\prod_{f\neq f_{2,3,4}}B_f^2  + \prod_{f\neq f_{1,2,3,4}}B_f^2 \left(I+ A_{v_0}\right) \Biggr) \\ \times \frac{2}{2^{|v|} 2^{|f|-4}}
     + \text{ traceless terms }.
\end{multline}
Taking the trace of this expression we obtain
\begin{align}
    {\rm Tr}(\mathcal{P}_4) &= \frac{2}{2^{|v|} 2^{|f|-4}} \left(2^{|e|} -4\times 2^{|e|} + 4 \times 3\times 2^{|e|-1} + 2\times 2^{|e|} -4 \times 3^2 \times 2^{|e| -2} +3^4 \times 2^{|e|-4} + 2^{|e|-4}\right)\\
    {\rm Tr}(\mathcal{P}_4)&= 9\times 2^{2g},
\end{align}
This $9$ fold degeneracy can be accounted for in terms of operators as described in the main text.

\bibliographystyle{apsrev4-1}
\bibliography{toric.bib}

\end{document}